\documentclass[acmsmall, screen]{acmart} 

\AtBeginDocument{
  }

\setcopyright{acmlicensed}
\copyrightyear{2026}
\acmYear{2026}
\acmDOI{XXXXXXX.XXXXXXX}

\acmJournal{TOMM}
\acmVolume{XX}
\acmNumber{XX}
\acmArticle{XX}
\acmMonth{1}

\usepackage{multirow}         
\usepackage{float}         
\usepackage{xcolor}        
     
\usepackage{booktabs}     
     
\usepackage{array}         
  
\usepackage{enumitem} 
\usepackage{amsmath}       
\usepackage[linesnumbered,ruled,vlined]{algorithm2e}    
\usepackage{subcaption}

\AfterEndPreamble{
  \hypersetup{
    colorlinks=true,       
    linkcolor=blue,         
    citecolor=blue,        
    urlcolor=blue,          
    unicode=true,          
    pdfborder={0 0 0}       
  }
}

\usepackage{cleveref} 

\Crefformat{section}{Section \textcolor{blue}{#2#1#3}}
\Crefformat{figure}{Figure \textcolor{blue}{#2#1#3}}
\Crefformat{subfigure}{Figure \textcolor{blue}{#2#1#3}}  
\Crefformat{table}{Table \textcolor{blue}{#2#1#3}}
\Crefformat{equation}{Equation \textcolor{blue}{#2(#1)#3}}

\crefformat{cite}{\textcolor{blue}{#2#1#3}}               
\Crefformat{cite}{Reference \textcolor{blue}{#2#1#3}}     
\crefformat{cites}{\textcolor{blue}{#2#1#3}}              
\Crefformat{cites}{References \textcolor{blue}{#2#1#3}}

\setlist{
    topsep=2pt,     
    partopsep=0pt,  
    itemsep=1pt,   
    parsep=0pt    
}

\begin{document}

\title{Cross-Modal Attention Network with Dual Graph Learning in Multimodal Recommendation}

\author{Ji Dai}
\email{daiji@bupt.edu.cn}
\affiliation{
  \institution{Beijing University of Posts and Telecommunications}
  \country{China}
}

\author{Quan Fang}
\authornote{Corresponding author: Quan Fang.}
\email{qfang@bupt.edu.cn}
\affiliation{
 \institution{Beijing University of Posts and Telecommunications}
 \city{Beijing}
     \country{China}}

\author{Jun Hu}
\email{jun.hu@nus.edu.sg}
\affiliation{
 \institution{National University of Singapore}
 \city{Singapore}
 \country{Singapore}}

\author{DeSheng Cai}
\email{caidsml@gmail.com}
\affiliation{
  \institution{Tianjin University of Technology}
  \country{China}
}

\author{Yang Yang}
\email{buaayangyang@buaa.edu.cn}
\affiliation{
 \institution{Beihang University}
 \city{Beijing}
\country{China}}
\affiliation{
 \institution{State Key Laboratory of CNS/ATM}
 \city{Beijing}
\country{China}}

\author{Can Zhao}
\email{zhaoc@adcc.com.cn}
\affiliation{
 \institution{Aviation Data Communication Corporation}
 \city{Beijing}
\country{China}}

\renewcommand{\shortauthors}{Ji Dai et al.}

\begin{abstract}
  Multimedia recommendation systems leverage user-item interactions and multimodal information to capture user preferences, enabling more accurate and personalized recommendations. Despite notable advancements, existing approaches still face two critical limitations: first, shallow modality fusion often relies on simple concatenation, failing to exploit rich synergic intra- and inter-modal relationships; second, asymmetric feature treatment---where users are only characterized by interaction IDs while items benefit from rich multimodal content---hinders the learning of a shared semantic space. To address these issues, we propose a \textbf{C}ross-modal \textbf{R}ecursive \textbf{A}ttention \textbf{N}etwork with dual graph \textbf{E}mbedding (\textbf{CRANE}). To tackle shallow fusion, we design a core \textbf{Recursive Cross-Modal Attention (RCA)} mechanism that iteratively refines modality features based on cross-correlations in a joint latent space, effectively capturing high-order intra- and inter-modal dependencies. For symmetric multimodal learning, we explicitly construct users' multimodal profiles by aggregating features of their interacted items. Furthermore, CRANE integrates a symmetric dual-graph framework---comprising a heterogeneous user-item interaction graph and a homogeneous item-item semantic graph---unified by a self-supervised contrastive learning objective to fuse behavioral and semantic signals. Despite these complex modeling capabilities, CRANE maintains high computational efficiency. Theoretical and empirical analyses confirm its scalability and high practical efficiency, achieving faster convergence on small datasets and superior performance ceilings on large-scale ones. Comprehensive experiments on four public real-world datasets validate an average 5\% improvement in key metrics over state-of-the-art baselines. Our code is publicly available at \url{https://github.com/MKC-Lab/CRANE}.
\end{abstract}

\ccsdesc[500]{Information systems~Recommender systems}
\ccsdesc[500]{Information systems~Multimedia information systems}
\ccsdesc[300]{Computing methodologies~Neural networks}

\keywords{Multimedia recommendation, Graph Neural Network, Multimodal Fusion}

\received{24 July 2025}
\received[revised]{05 December 2025}
\received[accepted]{03 January 2026}

\maketitle

\section{Introduction}

\begin{figure}[!t]
 \centering
 \includegraphics[width=\linewidth]{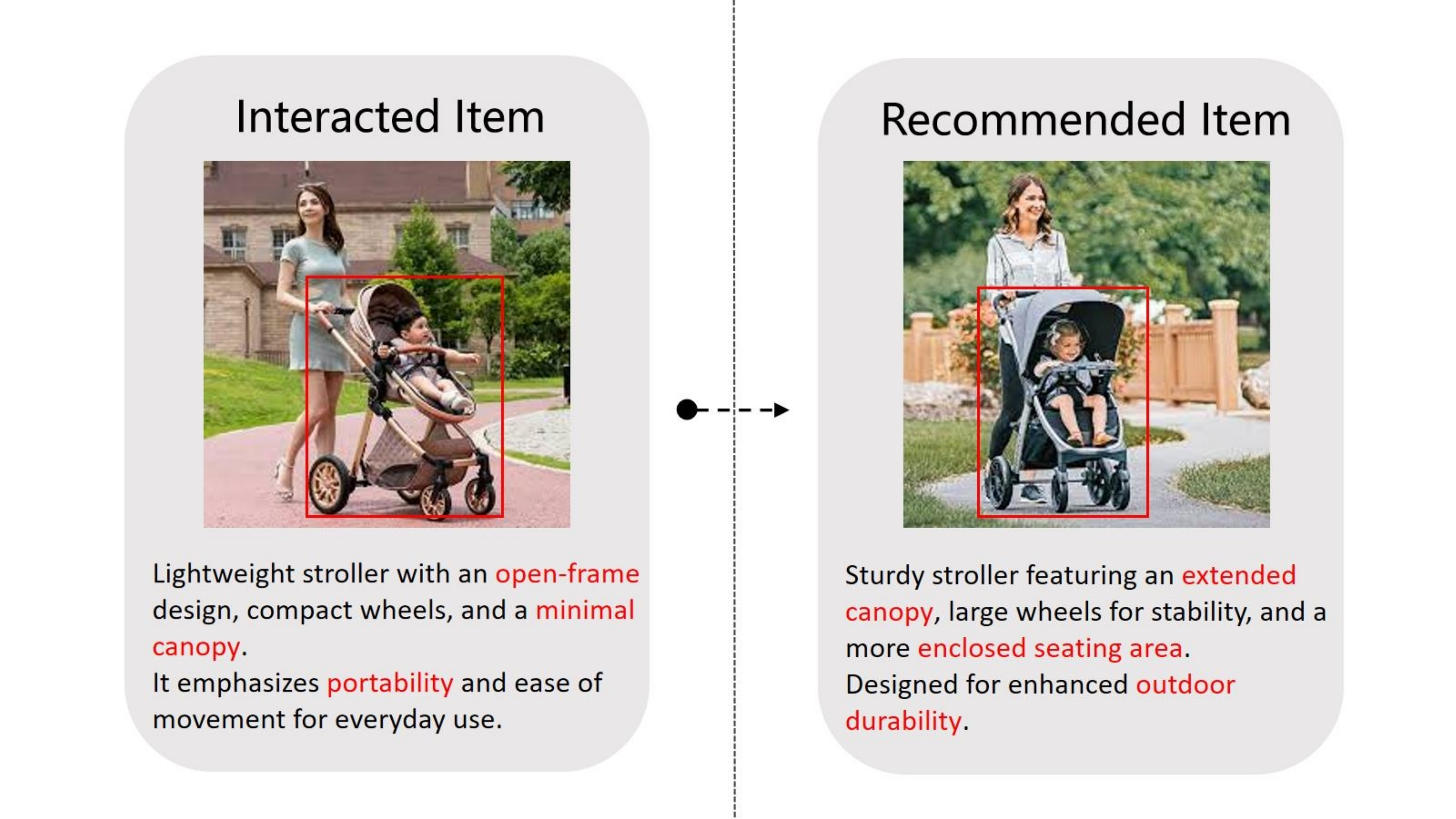}
 \caption{An illustration of how visual and textual modalities provide complementary signals to bridge the semantic gap. The system identifies latent user preferences by aligning fine-grained attributes, such as canopy design and outdoor durability, between the historical interaction and candidate recommendations.}
 \label{fig:motivation}
 \vspace{-2pt}
\end{figure}

With the rapid advancement of big data technology, Recommender Systems (RSs) have become indispensable instruments for online service platforms—ranging from e-commerce to social media—to mitigate information overload and deliver personalized user experiences~\cite{Wang2019NeuralGC,Chen2020BiasAD}. While traditional Collaborative Filtering (CF) has achieved remarkable success~\cite{HeM16,DBLP:conf/sigir/0001DWLZ020}, its efficacy is often severely constrained by inherent limitations such as data sparsity and the cold-start challenge~\cite{Li2023TransferableFF}, which hinder the capture of high-quality user preferences from limited behavioral logs~\cite{Zhou2023ACS}. To transcend these limitations, Multimodal Recommendation Systems (MRS) have emerged as a critical research paradigm~\cite{Liu2023MultimodalRS}. Fundamentally, MRS extends traditional ID-based paradigms by taking both historical user-item interactions and rich item-side multimodal features (e.g., product images, textual descriptions, and audio tracks) as joint inputs to generate accurate personalized item rankings as outputs. Unlike conventional approaches that treat items as atomic IDs, MRSs explicitly exploit this rich auxiliary media content to bridge the semantic gap between low-level interaction data and high-level user intent~\cite{Yuan2023WhereTG}. For instance, visual features capture aesthetic and structural details such as product design, while textual descriptions convey functional and contextual nuances like material composition (as shown in \Cref{fig:motivation}). Jointly modeling these heterogeneous modalities not only uncovers latent user interests that are not explicitly reflected in sparse interaction records~\cite{Chen2019PersonalizedFR} but also enhances system robustness and recommendation interpretability~\cite{Wei2023LightGTAL}.

Early works in this field, represented by VBPR~\cite{HeM16}, utilized deep learning to integrate multimodal features as supplementary side information. With the advent of Graph Neural Networks (GNNs), methods including MMGCN~\cite{Wei2019MMGCNMG}, GRCN~\cite{Wei2020GraphRefinedCN}, and DualGNN~\cite{Wang2023DualGNNDG} have achieved state-of-the-art performance by embedding high-order semantics into user and item representations. Notably, LATTICE~\cite{Zhang00WWW21} pioneered a dual-graph framework, constructing both a user-item interaction graph and a modality-based item similarity graph, which FREEDOM~\cite{DBLP:conf/mm/ZhouS23} further improved via graph freezing and denoising strategies.  From the perspective of fine-grained attributes, DIARec\cite{Vaghari2024DiarecDI} utilizes a hierarchical attention network to explicitly model the dynamic item-attribute interplay, thereby capturing multifaceted user intentions.

Despite these significant advancements, current research on MRS\cite{Liu2024MultimodalPA} continues to face two critical limitations. 
First, Shallow Modality Fusion: Existing studies often combine modality features using simple linear combinations or concatenation, treating all modalities equally. This approach overlooks complex, synergic intra- and inter-modal relationships.
Second, Asymmetric Representation: Most existing works model items with rich multimodal features while representing users solely through historical interactions, resulting in an item-centric bias. This asymmetry hinders the learning of a shared semantic space. 
Furthermore, recent extensions often introduce increased architectural complexity and computation costs~\cite{XuCWHKN25}, leaving the trade-off between performance and efficiency insufficiently addressed~\cite{DBLP:conf/www/ZhouZLZMWYJ23}.

To address these challenges, we introduce an innovative \underline{\textbf{C}}ross-modal \underline{\textbf{R}}ecursive \underline{\textbf{A}}ttention \underline{\textbf{N}}etwork with dual graph \underline{\textbf{E}}mbedding\footnote{While the full name derives from "Embedding", we use the acronym \textbf{CRANE} for brevity.}, \textbf{CRANE}. Illustrated in \Cref{fig:CRANE}, CRANE comprises three key components. First, we construct a heterogeneous user-item interaction graph using conventional collaborative signals to enable structural ID-based propagation. Second, to overcome fusion limitations, we propose a \textbf{Recursive Cross-Modal Attention (RCA)} mechanism. This module iteratively refines intra- and inter-modal correlations in a joint space, enabling dynamic weighting of modalities aligned with user preferences. Lastly, we construct a homogeneous item-item similarity graph based on fused embeddings, capturing fine-grained semantic affinities to complement the collaborative signal. The final representations jointly encode behavioral and semantic knowledge, yielding robust recommendations.

Notably, CRANE achieves these improvements without incurring high computational costs. Unlike methods with complex multi-stage architectures or heavy pre-training dependencies, CRANE leverages a lightweight, end-to-end design. Our theoretical and empirical analyses reveal that CRANE maintains a near-linear computational complexity, converging faster on small datasets and achieving higher performance ceilings on large ones with only marginal overhead.

Our main contributions are summarized as follows:
\begin{itemize}
    \item We propose the recursive cross-modal attention (RCA) mechanism that enables deep fusion of visual and textual features through iterative refinement in distinct latent spaces.
    \item We introduce a Symmetric Dual-Graph Framework, where we explicitly construct user multimodal profiles and integrate them with an item-item similarity graph, achieving balanced representation learning for both entities.
    \item CRANE achieves a significant performance improvement of 5\% on average across four real-world datasets compared to SOTA. Rigorous ablation studies confirm the model's effectiveness in capturing high-order cross-modal interactions.
    \item CRANE offers competitive efficiency, converging faster on simpler datasets and delivering more expressive representations on complex datasets  with manageable cost.
\end{itemize}

\section{Related Work}

\subsection{Multimodal Recommendation}
Multimodal recommendation models integrate visual, textual and behavioral data into collaborative filtering (CF) frameworks using deep learning or graph-based techniques. Initial approaches like VBPR~\cite{HeM16} incorporated visual content via pre-trained CNNs to enrich item representations, yet this strategy may introduce noise into the learned embeddings. Addressing the pervasive noise bottleneck, recent studies have introduced advanced filtering strategies: GCORec~\cite{vaghari2025group} utilizes a group-wise enhancement mechanism to dynamically amplify informative attributes, while DDRec~\cite{ping2025ddrec} pioneers a ``hard-and-soft" dual denoising paradigm that rigorously couples structural pruning with semantic feature purification. Attention mechanisms, such as those used in VECF~\cite{Chen2019PersonalizedFR} and MAML~\cite{DBLP:conf/mm/LiuCSWNK19}, focus on relevant multimodal information by segmenting images and modeling preferences for specific features. More recently, approaches like HAN~\cite{vaghari2025han} have further advanced this direction by employing hierarchical attention networks to capture latent context-aware preferences and dynamic attribute-item relationships.
Graph Neural Networks (GNNs) also advance multimodal recommendation by embedding high-order semantics into user and item representations. MMGCN~\cite{Wei2019MMGCNMG} constructs modality-specific user-item graphs, while GRCN~\cite{Wei2020GraphRefinedCN} refines interaction graphs to reduce noise. However, these methods can be computationally intensive, especially on large datasets.
Recent developments include DualGNN~\cite{Wang2023DualGNNDG}, which captures evolving user preferences across modalities, and LATTICE~\cite{Zhang00WWW21}, which dynamically updates item-item relationship graphs for each modality. Based on the Lattice, subsequent methods have further advanced graph structure learning ~\cite{Guo2023LGMRecLA}. In addition, the use of self-supervised learning~\cite{Wang2019NeuralGC} in multimodal recommendation offers new avenues for improving efficiency.

\subsection{Graph Convolution Network}
Graph Convolution Networks (GCNs) have gained popularity among researchers for their ability to effectively extract features from user behavior data, such as clicks or purchases, which could easily be modeled as an interaction bipartite graph~\cite{DBLP:conf/aaai/ChenWHZW20}. In particular, the NGCF model ~\cite{Wang2019NeuralGC} improves the capture of user behavior features by iteratively aggregating information from neighboring nodes in the user-item graph. Meanwhile, LightGCN \cite{DBLP:conf/sigir/0001DWLZ020} streamlines the traditional graph convolution process, optimizing it for recommendation tasks. Based on these advancements, GCN-based approaches have been expanded to multimedia recommendation systems ~\cite{Wang2023DualGNNDG}. For example, MMGCN ~\cite{Wei2019MMGCNMG} uses multiple GCN modules to handle various modalities, combining the resulting features to form complete item representations. Despite these innovations, propagation of modality noise throughout the graph remains a challenge due to the message-passing mechanism inherent in GCNs. Unlike methods that directly combine modality features, GRCN \cite{Wei2020GraphRefinedCN} incorporates a module to uncover latent relationships using graph-based techniques. However, excluding modality information may hinder the full exploration of user preferences.

\subsection{Multimodal Fusion} 
\label{sec:multimodal-fusion}
Currently, multimodal fusion strategies are commonly categorized into three types: early fusion, late fusion, and hybrid fusion~\cite{DBLP:journals/pami/BaltrusaitisAM19}.  Early fusion involves the amalgamation of extracted features in the initial stage. For example, methods like ACNet~\cite{DBLP:conf/icip/HuYFW19} utilize attention mechanisms to integrate data early in the process. In contrast, late fusion delays integration until each modality has undergone its own specific calculation and decision-making process (such as classification and regression task ). An example of this is NMCL \cite{DBLP:journals/tip/WeiWGNLC20}, which employs cooperative networks to enhance features for each modality using attention mechanisms, subsequently integrating predictions across modalities. Hybrid fusion merges the processes of early and late fusion to leverage their respective strengths. CELFT \cite{DBLP:conf/icmcs/Wang0YXCS21} exemplifies this approach by combining these techniques to overcome individual limitations. Multimodal recommendation models typically apply techniques such as the mean function, the attentive summation function, or the max-pooling~\cite{Hu2025NTSFormerAS,Zhang00WWW21}.

\section{Methodology}
\label{sec:methods}

\subsection{Problem Statement}

We formulate the multimodal recommendation task using the user-item interaction matrix $\mathbf{A}$ and item multimodal feature matrices ($\mathbf{X}^v, \mathbf{X}^t$) as inputs, with the objective of generating accurate top-$K$ recommendations. Detailed notations are provided in \Cref{tab:notations}.

\begin{figure*}[!t]
\centering
\includegraphics[width=\linewidth]{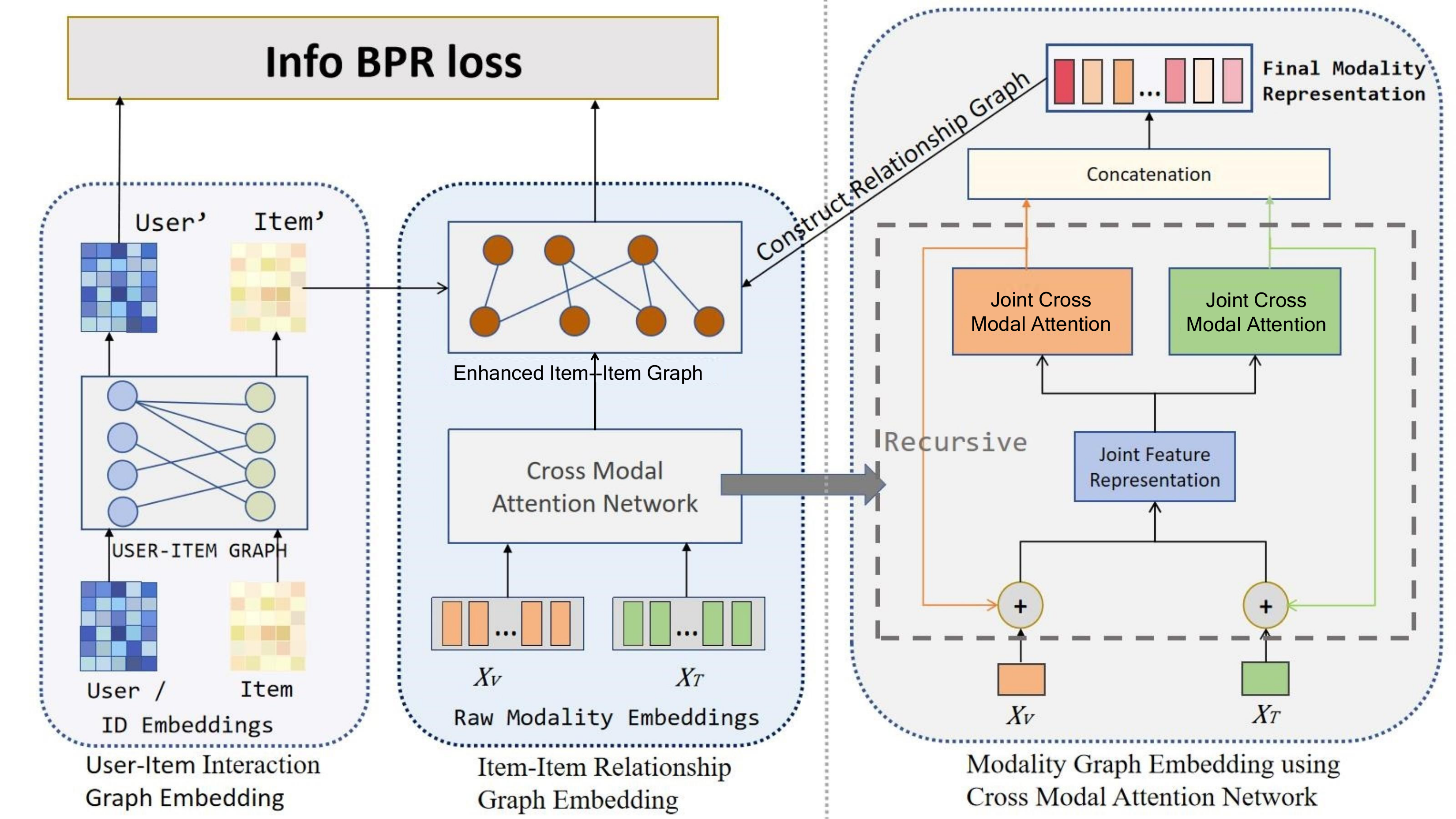}
\caption{The overall framework of CRANE. Initially, the model leverages the heterogeneous user-item graph (left) to extract fundamental collaborative signals from historical interactions. Afterwards, the core RCA mechanism (right) iteratively aligns and refines visual and textual features to uncover deep intra- and inter-modal dependencies. These refined features are then employed to construct a homogeneous item-item semantic graph (middle), injecting high-order semantic affinities into item representations. Finally, the collaborative and semantic views are integrated to predict user preferences, optimized jointly by BPR and contrastive losses.}
\label{fig:CRANE}
\end{figure*}

As illustrated in \Cref{fig:CRANE}, the proposed CRANE framework integrates collaborative and semantic information through three progressive phases: we first utilize a heterogeneous user-item graph $\mathcal{G}_{\text{UI}}$ to extract fundamental behavioral patterns; subsequently, we employ the recursive cross-modal attention (RCA) mechanism to iteratively refine visual and textual semantics in a shared latent space; and finally, we construct a homogeneous item-item graph $\mathcal{G}_{\text{II}}$ to incorporate high-order semantic affinities, fusing them with collaborative signals for the final prediction.

\begin{table}[t]

  \caption{Summary of Key Notations}
  \label{tab:notations}
  \begin{tabular}{ll}
    \toprule
    \textbf{Notation} & \textbf{Description} \\
    \midrule
    \multicolumn{2}{l}{\textit{\textbf{Sets and Graph Structures}}} \\
    $\mathcal{U}, \mathcal{I}$ & Sets of users and items; total numbers $M = |\mathcal{U}|, N = |\mathcal{I}|$. \\
    $\mathcal{G}_{\text{UI}}, \mathcal{G}_{\text{II}}$ & Heterogeneous user-item graph and homogeneous item-item graph. \\
    $\mathbf{A}$ & User-item interaction matrix, $\mathbf{A} \in \{0,1\}^{M \times N}$. \\
    $\mathbf{S}$ & Item-item semantic similarity matrix, $\mathbf{S} \in \mathbb{R}^{N \times N}$. \\
    $\mathcal{N}(u), \mathcal{N}_S(i)$ & Interaction neighbors for user $u$, and semantic neighbors for item $i$. \\
    \midrule
    \multicolumn{2}{l}{\textit{\textbf{Embeddings and Features}}} \\
    $\mathbf{x}_i^v, \mathbf{x}_i^t$ & Raw visual ($d_v$-dim) and textual ($d_t$-dim) feature vectors for item $i$. \\
    $\mathbf{E}$ & Unified multimodal embedding matrix, $\mathbf{E} \in \mathbb{R}^{(M+N) \times (d_v+d_t)}$. \\
    $\mathbf{e}_u, \mathbf{e}_i$ & Collaborative embeddings learned from $\mathcal{G}_{\text{UI}}$. \\
    $\mathbf{h}_u, \mathbf{h}_i$ & Semantic-aware embeddings learned from $\mathcal{G}_{\text{II}}$. \\
    $\mathbf{z}_u, \mathbf{z}_i$ & Final unified representations for prediction. \\
    \midrule
    \multicolumn{2}{l}{\textit{\textbf{Model Parameters}}} \\
    $L_{UI}, L_{II}$ & Number of GCN layers for $\mathcal{G}_{\text{UI}}$ and $\mathcal{G}_{\text{II}}$, respectively. \\
    $R, k$ & RCA iteration depth ($R$) and neighbor count for $\mathcal{G}_{\text{II}}$ ($k$). \\
    $\mathcal{L}_{CL}$ & Contrastive alignment loss between collaborative and semantic views. \\
    $\beta, \lambda, \tau$ & Loss weights for contrastive task and regularization; temperature $\tau$. \\
    \bottomrule
  \end{tabular}
  \vspace{4pt}
\end{table}

\subsection{User-Item Collaborative Embedding via Heterogeneous Graph Convolution}
\label{subsec:uic-embedding}
In this module, we aim to learn collaborative behavior embeddings for users and items, leveraging only user-item interaction records. The process involves constructing an interaction graph and propagating information via heterogeneous graph convolution.

\subsubsection{User-Item Interaction Graph Construction.}
We define the user-item interaction matrix $\mathbf{A} \in \mathbb{R}^{M \times N}$ :
\begin{equation}
    A_{ui} = 
    \begin{cases} 
    1, & \text{if user } u \text{ interacted with item } i \\
    0, & \text{otherwise}
    \end{cases}
\end{equation}
This matrix forms a bipartite graph $\mathcal{G}_{\text{UI}}$. To alleviate overfitting, we adopt DropEdge to randomly drop edges.  Instead of a fixed rate, we calculate an adaptive retention probability $P_{ui}$ based on structural connectivity:
\begin{equation}
    P_{ui} = \frac{1}{\sqrt{d_u \cdot d_i}} \tag{2}
\end{equation}
where $d_u$ and $d_i$ denote the degrees of user $u$ and item $i$, respectively. This design ensures that edges connected to high-degree nodes are more likely to be dropped, preserving structural diversity. The sampled adjacency matrix $\tilde{\mathbf{A}}$ is generated as:
\begin{equation}
    \tilde{A}_{ui} = \text{Sample}(A_{ui}, P_{ui}) = 
    \begin{cases} 
    1, & \text{with probability } P_{ui} \\
    0, & \text{with probability } 1 - P_{ui}
    \end{cases}
\end{equation}

\subsubsection{Heterogeneous Graph Convolution.}
We perform heterogeneous graph convolution on the sampled graph $\tilde{\mathbf{A}}$ to aggregate neighborhood information. Let $\mathbf{e}_u^{(l)}$ and $\mathbf{e}_i^{(l)}$ denote the embeddings of user $u$ and item $i$ at layer $l$. The message propagation rule is defined as:
\begin{equation}
    \mathbf{e}_u^{(l+1)} = \sigma \left( \sum_{i \in \mathcal{N}(u)} \tilde{A}_{ui} \cdot \frac{1}{\sqrt{d_u d_i}} \mathbf{W}^{(l)} \mathbf{e}_i^{(l)} \right)
    \label{equ:equ4}
\end{equation}
\begin{equation}
    \mathbf{e}_i^{(l+1)} = \sigma \left( \sum_{u \in \mathcal{N}(i)} \tilde{A}_{ui} \cdot \frac{1}{\sqrt{d_i d_u}} \mathbf{W}^{(l)} \mathbf{e}_u^{(l)} \right)
    \label{equ:equ5}
\end{equation}
where $\mathcal{N}(u)$ and $\mathcal{N}(i)$ are the neighbor sets in the interaction graph. The term $\tilde{A}_{ui}$ acts as a structural filter based on the probability $P_{ui}$, while the fraction term serves as the standard Laplacian normalization, and $\mathbf{W}^{(l)}$ is the trainable weight matrix.

\subsection{Comprehensive Multi-Modality Embedding via Recursive Cross-Modal Attention}
\label{subsec:rca}

In this part, our objective is to construct unified multimodal embeddings for users and items that can effectively integrate visual and textual semantics. The process begins by forming modality-aware user representations based on their historical interactions and culminates in a recursive cross-modal attention (RCA) module that iteratively aligns and refines multimodal information.

\subsubsection{User Modality Construction.}

Inspired by CAMR~\cite{wang2025towards}, which constructs user modality embeddings via graph convolution using normalized mean pooling, we construct user profiles by aggregating the multimodal features of interacted items.

Based on the interaction set $\mathcal{N}(u) = \{i \mid A_{ui} = 1\}$, and given the pre-extracted visual and textual feature vectors $\mathbf{x}_i^v \in \mathbb{R}^{d_v}$ and $\mathbf{x}_i^t \in \mathbb{R}^{d_t}$ for item $i$, the modality-aware user profile is obtained by:

\begin{equation}
    \mathbf{x}_u^m = \sum_{i \in \mathcal{N}(u)} \mathbf{x}_i^m, \quad m \in \{v, t\}
\label{equ:user}
\end{equation}

Distinct from CAMR, our approach eschews averaging to avoid diluting the collaborative signal inherent in binary implicit feedback. We provide a detailed empirical justification for this aggregation strategy in \Cref{sec:user-profile}.

Stacking the vectors of all users and items yields two unified modality matrices:
\begin{equation}
    \mathbf{X}^m = 
    \begin{bmatrix}
    \mathbf{X}_{\mathcal{U}}^m \\
    \mathbf{X}_{\mathcal{I}}^m
    \end{bmatrix}, 
    \quad m \in \{v, t\}
\end{equation}
where $\mathbf{X}_{\mathcal{U}}^m$ and $\mathbf{X}_{\mathcal{I}}^m$ represent the feature matrices for the user set and item set, respectively. These unified matrices place all entities into shared visual and textual spaces, serving as the basis for our recursive fusion.

\subsubsection{Recursive Cross-Modal Attention (RCA).}
To obtain comprehensive multimodal embeddings, we propose the RCA mechanism that progressively enhances the alignment between modalities. The overall procedure of the RCA mechanism is summarized in Algorithm~\ref{alg:rca}. Each iteration consists of three conceptual steps: (1) joint feature integration, (2) modality-specific correlation learning, and (3) feature refinement.

\begin{algorithm}[t]
\caption{Recursive Cross-Modal Attention (RCA)}
\label{alg:rca}
\SetAlgoLined 
\DontPrintSemicolon

\SetKwInOut{Input}{Require}
\SetKwInOut{Output}{Ensure}

\newcommand{\atcp}[1]{\tcp*[r]{#1}} 
\SetKwComment{tcp}{ // }{} 

\Input{Initial modality matrices $\mathbf{X}^{v(0)}, \mathbf{X}^{t(0)}$; Iteration depth $R$; \\
       Trainable parameters $\Theta = \{\mathbf{W}_{tr}, \mathbf{b}_{tr}, \mathbf{W}_m, \mathbf{W}_a^m, \mathbf{W}_f^m\}_{m \in \{v,t\}}$.}
\Output{Comprehensive multimodal embedding $\mathbf{X}_{att}$.}

\BlankLine
// \textit{Phase 1: Recursive Refinement Process}\;
\For{$r = 1, \cdots, R$}{
    \BlankLine
    // \textit{Step 1: Joint Feature Integration}\;
    $\mathbf{E} \leftarrow [\mathbf{X}^{v(r-1)}; \mathbf{X}^{t(r-1)}]$ \tcp*{Concatenate modalities}
    $\mathbf{E}' \leftarrow \mathbf{E}\mathbf{W}_{tr} + \mathbf{b}_{tr}$ \tcp*{Transform to joint latent space}
    
    \For{\text{each modality} $m \in \{v, t\}$}{
        \BlankLine
        // \textit{Step 2 \& 3: Cross-Modal Attention \& Update}\;
        $\mathbf{P}^m \leftarrow \mathbf{E}'\mathbf{W}_m$ \tcp*{Project to modality anchors}
        $\mathbf{C}^m \leftarrow \tanh(\mathbf{X}^{m(r-1)} (\mathbf{P}^m)^\top)$ \tcp*{Compute Correlation Matrix}
        $\mathbf{F}^m \leftarrow \text{ReLU}(\mathbf{C}^m \mathbf{X}^{m(r-1)} \mathbf{W}_a^m)$ \tcp*{Refine features}
        $\mathbf{X}^{m(r)} \leftarrow \mathbf{F}^m + \mathbf{X}^{m(r-1)} \mathbf{W}_f^m$ \tcp*{Update with residual connection}
    }
}

\BlankLine
// \textit{Phase 2: Final Representation}\;
$\mathbf{X}_{att} \leftarrow [\mathbf{X}^{v(R)}; \mathbf{X}^{t(R)}]$\;
\Return $\mathbf{X}_{att}$

\vspace{7pt}
\end{algorithm}

\paragraph{Step 1: Joint feature integration.}
We begin by concatenating the visual and textual embeddings for each entity to form a joint embedding matrix $\mathbf{E}$:
\begin{equation}
    \mathbf{E} = [\mathbf{X}^v ; \mathbf{X}^t] \in \mathbb{R}^{(M+N) \times (d_v + d_t)}
\end{equation}
A transformation layer is then applied to mix complementary information from both modalities:
\begin{equation}
    \mathbf{E}' = \mathbf{E}\mathbf{W}_{tr} + \mathbf{b}_{tr}
\end{equation}
where $\mathbf{W}_{tr}$ and $\mathbf{b}_{tr}$ are learnable parameters. To relate the joint embedding back to each modality, we project the transformed features into modality-specific latent spaces:
\begin{equation}
    \mathbf{P}^m = \mathbf{E}'\mathbf{W}_m, \quad m \in \{v, t\}
\end{equation}
where $\mathbf{W}_m$ aligns joint information with modality $m$. These projected features $\mathbf{P}^m$ serve as anchors to evaluate relevance between the joint representation and original modality features.
\paragraph{Step 2: Cross-modal attention mapping.}
To assess how each modality contributes to the joint semantic space, we compute the cross-modal correlation matrices $\mathbf{C}^m$:
\begin{equation}
    \mathbf{C}^m = \tanh(\mathbf{X}^m (\mathbf{P}^m)^\top), \quad m \in \{v, t\}
\end{equation}
Here, $\mathbf{C}^m \in \mathbb{R}^{(M+N) \times (M+N)}$ quantifies how strongly the entities' modality-$m$ features relate to the aggregated multimodal representation. In this way, RCA identifies modality-specific signals most relevant to the joint semantics.

\paragraph{Step 3: Feature refinement and update.}
Each modality feature matrix is refined by weighting entity-level interactions via the learned correlation matrix $\mathbf{C}^m$:
\begin{equation}
    \mathbf{F}^m = \text{ReLU}(\mathbf{C}^m \mathbf{X}^m \mathbf{W}_a^m)
\end{equation}
where $\mathbf{W}_a^m$ extracts informative cross-modal patterns. To preserve the original modality structure while injecting attention-enhanced information, we combine the refined and original features via a residual connection:
\begin{equation}
    (\mathbf{X}^m)' = \mathbf{F}^m + \mathbf{X}^m \mathbf{W}_f^m, \quad m \in \{v, t\}
\end{equation}
where $\mathbf{W}_f^m$ is a learnable transformation matrix that ensures semantic alignment between the raw and refined features. The updated matrices $(\mathbf{X}^v)'$ and $(\mathbf{X}^t)'$ are then used as the inputs for the next iteration.

\paragraph{Recursive refinement.}
The three steps above are repeated for $R$ iterations, enabling the model to capture increasingly higher-order multimodal dependencies. After recursion, the final multimodal embedding is obtained by feature concatenation:
\begin{equation}
\label{eq:atten}
    \mathbf{X}_{att} = [\mathbf{X}^{v(R)} ; \mathbf{X}^{t(R)}]
\end{equation}
This comprehensive representation $\mathbf{X}_{att}$ effectively encodes deep intra- and inter-modal semantics, serving as a robust initialization for constructing the subsequent homogeneous item-item graph.

\subsection{Homogeneous Graph Convolution and Final Representation}

 \begin{figure*}[!t]
  \centering
  \includegraphics[width=\linewidth]{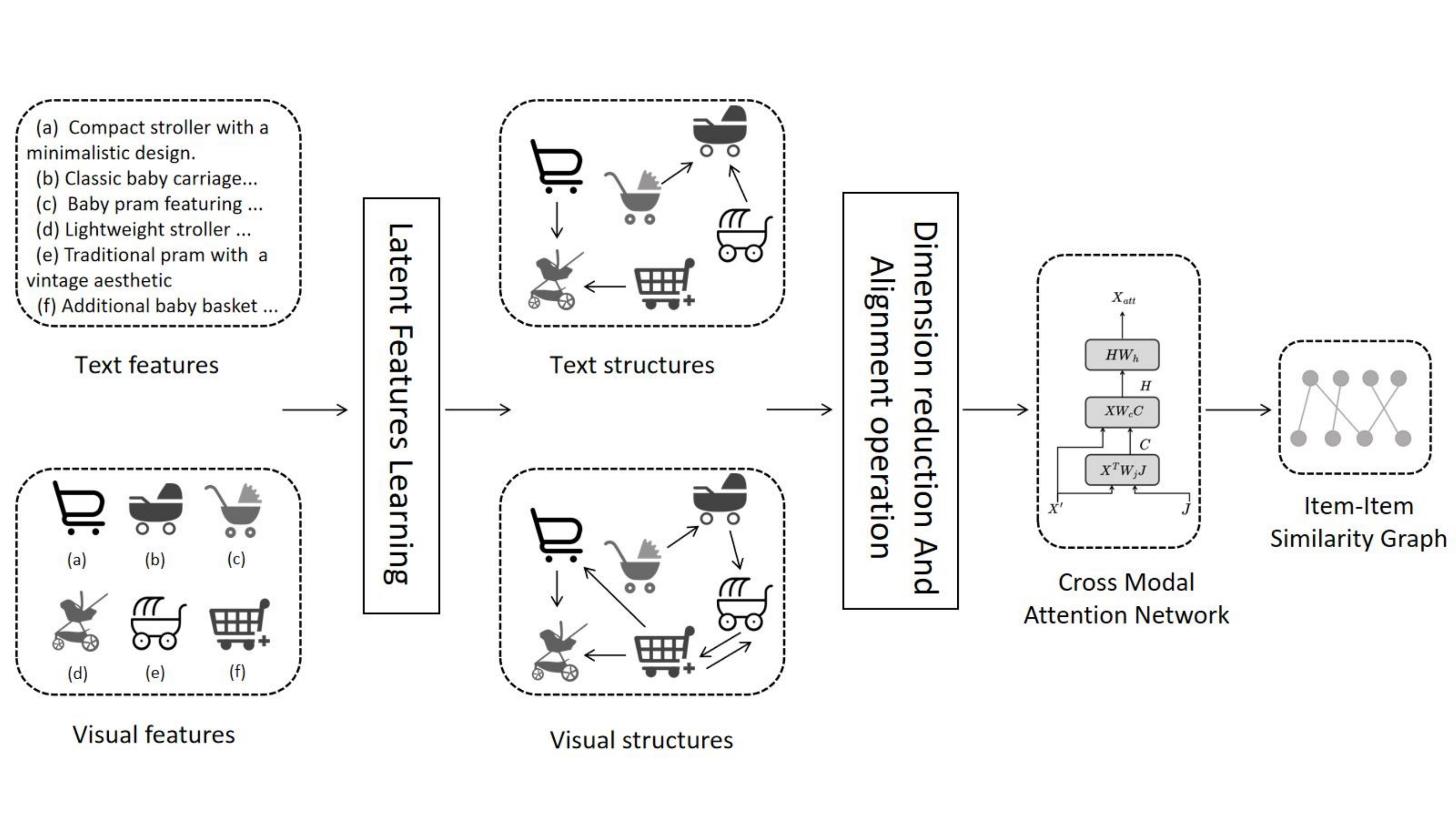}
  \caption{The pipeline of Item-Item Semantic Graph construction. It details the transformation from raw multimodal features to the final adjacency matrix. The process involves latent feature alignment, dimension reduction, and the application of the Cross-Modal Attention Network to derive sparse, high-order semantic connections between items.
 }
  \label{fig:process}
\end{figure*}

While the RCA mechanism captures fine-grained feature interactions, modeling the global semantic structure among items remains essential. The detailed construction pipeline is depicted in \Cref{fig:process}. To explicitly capture global semantic affinities, we construct a homogeneous item-item graph $\mathcal{G}_{II}$.  Based on the refined multimodal embeddings $X_{att}$ derived in \Cref{subsec:rca}, we first compute the pairwise cosine similarity matrix $S \in \mathbb{R}^{N \times N}$, where the entry $S_{ij}$ is calculated as:
\begin{equation}
    S_{ij} = \frac{x_{i, att}^\top x_{j, att}}{\|x_{i, att}\| \|x_{j, att}\|}
\end{equation}
Constructing a fully connected graph based on $S$ incurs cubic computational costs and introduces noise. Therefore, we employ a sparsification strategy to retain only the top-$K$ semantic neighbors for each item. Let $\mathcal{N}_S(i)$ denote the set of these neighbors. We obtain the sparse adjacency weights $\tilde{S}_{ij}$ by row-normalizing the retained entries in $S$, ensuring numerical stability for propagation.

With the semantic structure established, we employ graph convolution to propagate high-order signals. A critical design choice is the initialization strategy: instead of starting from random vectors, we utilize the collaborative embeddings $e_i$ learned from the user-item graph (\Cref{subsec:uic-embedding}) as the initial state, i.e., $h_i^{(0)} = e_i$. This facilitates the immediate fusion of collaborative and semantic views. The propagation rule at layer $l$ is formulated as:
\begin{equation}
    h_i^{(l+1)} = \sum_{j \in \mathcal{N}_S(i)} \tilde{S}_{ij} h_j^{(l)}
    \label{eq:ii_gcn}
\end{equation}
After stacking $L_{II}$ layers, we obtain the final semantic-aware item embedding $h_i = h_i^{(L_{II})}$.

To address the inherent feature asymmetry—where users lack explicit semantic attributes—we construct the user's semantic profile $h_u$ by aggregating the learned embeddings of their interacted items: 
\begin{equation}
h_u = \sum_{i \in \mathcal{N}(u)} h_i
\end{equation}
 Finally, we fuse the collaborative signal ($e$) and semantic signal ($h$) to form the unified representations $z_u$ and $z_i$:
\begin{equation}
    z_u = e_u + h_u, \quad z_i = e_i + h_i
    \label{eq:final_fusion}
\end{equation}
These unified representations serve as comprehensive profiles, encoding both high-order collaborative structure and refined multimodal semantics for the final prediction.

\subsection{Optimization and Prediction}
Based on the final unified representations, the predicted preference score $\hat{y}_{ui}$ for a user $u$ towards an item $i$ is calculated via the inner product: $\hat{y}_{ui} = \mathbf{z}_u^\top \mathbf{z}_i$. To optimize the model parameters, we design a multi-task objective function that integrates the primary recommendation task with a self-supervised auxiliary task.

For recommendation, we employ the Bayesian Personalized Ranking (BPR) loss. This objective is designed to enforce the correct relative order of items, assuming that observed interactions should be ranked higher than unobserved ones. The loss function is defined on pairwise training triplets $\mathcal{D} = \{(u, i, j) \mid (u, i) \in \mathcal{R}^+, (u, j) \in \mathcal{R}^-\}$:
\begin{equation}
    \mathcal{L}_{BPR} = \sum_{(u,i,j) \in \mathcal{D}} -\log \sigma(\hat{y}_{ui} - \hat{y}_{uj})
\end{equation}
Minimizing this loss effectively widens the margin between the predicted scores of positive item $i$ and negative item $j$, thereby optimizing the ranking performance.

However, relying solely on the supervised BPR signal may lead to suboptimal representations due to data sparsity. To mitigate this, we introduce a contrastive alignment mechanism ~\cite{DBLP:journals/kais/LiuYGYD24}. This auxiliary task treats the collaborative view ($\mathbf{e}$) and the semantic view ($\mathbf{h}$) of the same node as positive pairs, while treating embeddings from different nodes as negative pairs. For a randomly sampled user batch $\mathcal{B}$, the user-side InfoNCE loss is formulated as:
\begin{equation}
    \mathcal{L}_{CL}^U = \sum_{u \in \mathcal{B}} -\log \frac{\exp(\text{sim}(\mathbf{e}_u, \mathbf{h}_u)/\tau)}{\sum_{u' \in \mathcal{B}} \exp(\text{sim}(\mathbf{e}_u, \mathbf{h}_{u'})/\tau)}
    \label{equ:equ20}
\end{equation}
where $\tau$ is the temperature hyperparameter. Intuitively, optimizing this objective maximizes the mutual information between the two views, encouraging the model to learn consistent user representations across the behavioral and semantic spaces. A symmetric item-side loss, $\mathcal{L}_{CL}^I$, is computed in an analogous manner.

The final objective loss combines the recommendation loss, the contrastive alignment losses, and $L_2$ regularization:
\begin{equation}
    \mathcal{L} = \mathcal{L}_{BPR} + \beta(\mathcal{L}_{CL}^U + \mathcal{L}_{CL}^I) + \lambda \|\Theta\|_2^2
\end{equation}
where $\beta$ balances the self-supervised signal and $\lambda$ controls the regularization strength. By jointly optimizing these components, CRANE effectively fuses collaborative signals with multimodal semantics, ensuring robustness.

\section{Experiments}
\label{sec:experiments}

\subsection{Datasets}
\label{sec:datasets}

Following previous work\cite{Zhang00WWW21,DBLP:conf/mm/ZhouS23}, We evaluate our model on four public Amazon review datasets\cite{He2016UpsAD}: \textbf{Baby}, \textbf{Sports}, \textbf{Clothing}, and \textbf{Electronics}. These datasets are widely-used benchmarks that vary in domain and scale. Notably, the Electronics dataset (63,001 items) is included to validate the model's performance and scalability in scenarios with massive item pools.  To ensure fair comparison with existing MMRec works, we use the preprocessed data (including user-item interaction records and extracted multimodal embeddings) provided in the public  \href{https://drive.google.com/drive/folders/13cBy1EA_saTUuXxVllKgtfci2A09jyaG}{Google Drive repository}\footnotemark. For item multimodal features, we use the provided 4096-dimensional visual features (extracted by ResNet50) and 384-dimensional textual features (extracted by BERT). The statistics of the four datasets are summarized in ~\Cref{tab:dataset_stats}.

\footnotetext{\url{https://drive.google.com/drive/folders/13cBy1EA_saTUuXxVllKgtfci2A09jyaG} }

\noindent\begin{table}[!t]
\centering
\caption{Statistics of the Four Datasets}
\label{tab:dataset_stats}
\resizebox{\linewidth}{!}{
\begin{tabular}{lccccc}
\toprule
Dataset & \# Users & \# Items & \# Interactions & \# Sparsity& {\# Average Interactions per User}\\
\midrule
Baby & 19,445 & 7,050 & 160,792 & 99.88\% & {8.27}\\
Sports & 35,598 & 18,357 & 296,337 & 99.95\% & {8.32}\\
Clothing & 39,387 & 23,033 & 278,677 & 99.97\% & {7.07}\\
{Electronics}& {192,403}& {63,001}& {1,689,188}& {99.99\%}& {7.24}\\
\bottomrule
\end{tabular}
\vspace{6pt}
}
\end{table}

\subsection{Compared Baselines}

To evaluate the effectiveness of CRANE, we compare it against a comprehensive set of state-of-the-art methods, categorized into general collaborative filtering and multimodal recommendation models. Note that all baselines follow the rigorous comparison protocol detailed in \Cref{sec:exp_settings} to ensure fair evaluation.

\begin{itemize}
\item \textbf{BPR}~\cite{DBLP:journals/corr/abs-1205-2618} enhances latent representations of users and items in matrix factorization (MF) via BPR loss.
\item \textbf{LightGCN}~\cite{DBLP:conf/sigir/0001DWLZ020} streamlines recommendation by focusing on core structural information in user-item interactions, boosting collaborative filtering efficiency.
\item \textbf{VBPR}~\cite{HeM16} is a visual Bayesian personalized  model that integrates item image features to refine item ranking.
\item \textbf{MMGCN}~\cite{Wei2019MMGCNMG} is a multimodal graph convolutional network that fuses visual and textual features to enhance item representation learning.
\item \textbf{SLMRec}~\cite{Tao2023SelfSupervisedLF}  enhances multimedia recommendation via self-supervised learning with modal-agnostic (FD, FM) and modal-specific (FAC) tasks, capturing latent multi-modal patterns.
\item \textbf{LATTICE}~\cite{Zhang00WWW21} is a collaborative latent topic-aware filtering model that integrates topic modeling to capture semantic relationships between items and users.
\item \textbf{FREEDOM}~\cite{DBLP:conf/mm/ZhouS23}  freezes the pre-constructed item-item graph and denoises the user-item graph via degree-sensitive pruning, then fuses both with light-weighted GCN. 
\item \textbf{LGMRec}~\cite{Guo2023LGMRecLA} addresses coupling and locality issues by modeling local and global user interests via local graph embedding and global hypergraph learning.

\item \textbf{DGAVE}~\cite{DBLP:journals/tmm/ZhouM24} enhances interpretability via multimodal-to-text projection, frozen item-item graph, and disentangled VAE with mutual information maximization.
\item \textbf{LPIC}~\cite{DBLP:journals/tomccap/LiuSXWG25} models modality-specific user interests via learnable prompt embeddings (fusing item ID, text, visual features) and ID-guided contrastive learning. 

\end{itemize}

\subsection{Experimental Settings}

\label{sec:exp_settings}

To ensure fair comparison and reproducibility, our implementation is built upon MMRec~\cite{Zhou2023ACS} \footnotemark . We follow its standard protocols, while specific deviations are detailed below:

\footnotetext{\url{https://github.com/enoche/MMRec} }

\subsubsection{Reproducibility Protocol and Data Splitting.}
\label{sec:data_prep}

To ensure a rigorous evaluation, we apply a standardized data processing protocol consistent across all baselines. First, regarding preprocessing, we filter the original interaction matrix to retain only users with at least four interaction records and convert the 1-5 star ratings into binary implicit signals (1 for interaction, 0 otherwise), without constructing profiles based on interaction frequency. Second, for data splitting, the processed interactions are randomly divided into training, validation, and testing sets following a strict 8:1:1 ratio. To enable precise replication and robust statistical analysis, we generate five fixed data splits using specific random seeds (9, 672, 5368, 12784, 2023). All improvements reported for CRANE are statistically significant, confirmed by a paired t-test ($p < 0.005$).

\subsubsection{Hyperparameter Settings.}
\label{sec:implementation}
We implement CRANE using PyTorch and conduct all experiments on an NVIDIA RTX 3090 GPU. Our parameter configuration follows a structured strategy.  First, regarding model architecture, key structural hyperparameters—including graph layers ($L_{UI}, L_{II}$), recursion depth ($R$), and neighbor count ($k$)—are determined based on the sensitivity analysis in \Cref{sec:sensitivity}. Second, for optimization and loss, we perform a rigorous grid search on the validation set: the learning rate is searched in $\{10^{-3}, 5\times10^{-4}, 10^{-4}, 5\times10^{-5}\}$, the $L_2$ regularization $\lambda$ in $\{10^{-3}, 10^{-4}, 10^{-5}\}$, the contrastive temperature $\tau$ in $\{0.1, \dots, 1.0\}$, and the weight $\beta$ in $\{0.1, \dots, 0.9\}$. Third, for common settings, we adopt standard configurations proven effective in prior works \cite{DBLP:conf/mm/ZhouS23,Zhang00WWW21}, fixing the embedding dimension to 64 and batch size to 1024. We employ  Xavier initializer  and Adam optimizer across all models. Training is terminated if Recall@20 on the validation set does not improve for 5 successive epochs (Early Stopping). All final optimal values are listed in \Cref{tab:hyperparameters}.
\setcounter{table}{2}

\begin{table}[t]
  \centering
  
  \refstepcounter{table} 
  \label{tab:hyperparameters}
  
  \caption*{\centering Table \thetable. Hyperparameter settings for CRANE on four datasets.}
  
  \renewcommand{\arraystretch}{0.9} 
  
  \resizebox{0.85\linewidth}{!}{
    \setlength{\tabcolsep}{2.5mm} 
    \begin{tabular}{lcccc}
      \toprule
      \textbf{Hyperparameter} & \textbf{Baby} & \textbf{Sports} & \textbf{Clothing} & \textbf{Electronics} \\
      
      \midrule
      \multicolumn{5}{l}{\textbf{Model Architecture}} \\
      $L_{UI}$ (User-Item layers)          & 2 & 2 & 2 & 2 \\
      $L_{II}$ (Item-Item layers)          & 1 & 1 & 1 & 1 \\
      $R$ (Attention rounds)               & 3 & 2 & 3 & 3 \\
      $k$ (Neighbor count)                 & 15 & 15 & 10 & 10 \\
      
      \midrule
      \multicolumn{5}{l}{\textbf{Optimization \& Loss}} \\
      Learning rate                        & 1e-4 & 5e-5 & 1e-4 & 1e-4 \\
      $L_2$ regularization ($\lambda$)     & 1e-3 & 1e-3 & 1e-3 & 5e-4 \\
      Contrastive temp. ($\tau$)           & 0.6 & 0.5 & 0.4 & 0.5 \\
      Contrastive weight ($\beta$)         & 0.8 & 0.7 & 0.7 & 0.6 \\
      
      \midrule
      \multicolumn{5}{l}{\textbf{Common Settings}} \\
      Embedding size       & \multicolumn{4}{c}{64} \\
      Batch size           & \multicolumn{4}{c}{1024} \\
      Training epochs      & \multicolumn{4}{c}{250} \\
      Early stopping patience & \multicolumn{4}{c}{5} \\
      \bottomrule
    \end{tabular}
  }
  \vspace{-8.8pt}

\end{table}

\subsubsection{Evaluation Metrics and Sampling.} 
We utilize two widely accepted evaluation metrics, \textbf{Recall@K} and \textbf{NDCG@K}, to assess the performance of top-K recommendations, with $K$ set to 10 and 20. Recall@K measures the proportion of relevant items in the top-K list, while NDCG@K evaluates the ranking quality by assigning higher weights to relevant items ranked higher. For negative sampling during training, we adopt simple uniform random sampling across all datasets, pairing each positive interaction with one unobserved item.

\subsubsection{Baseline Comparison Protocol.}
\label{sec:baseline_protocol}
To ensure absolute fairness, we adhere to a strict evaluation protocol. (1) For Data Consistency: all baselines \textbf{are trained} and evaluated using the exact same preprocessed data and splits defined in Section~\ref{sec:data_prep}. (2) For models with publicly available code (e.g., FREEDOM, LATTICE), we re-implement them using official sources and \textbf{tune} hyperparameters to match their original papers. We verify that our reproduction results \textbf{are} consistent (difference $<0.5\%$), and thus report scores from the original publications to respect prior work. (3) Regarding the \textbf{Electronics} dataset, where official results \textbf{are} often unavailable, we report our own re-implemented results. Entries are marked with '–' if a model's code \textbf{is} not public or \textbf{fails} to execute on this larger scale.\\ \\

\begin{table*}[!h]
\scriptsize
\setlength{\tabcolsep}{0pt} 
\renewcommand{\arraystretch}{1.25} 
\centering
\caption{Overall Performance Comparison of different recommender systems. We \textbf{boldface} the best performance and \underline{underline} the second-best one.}
\label{tab:performance}

    \begin{tabular*}{\textwidth}{@{\extracolsep{\fill}} lc|cc|ccccccccc}
\hline
\multicolumn{2}{c|}{\multirow{2}{*}{}} & \multicolumn{2}{c|}{General Model} & \multicolumn{9}{c}{Multimodal Model} \\
Dataset & Metric & BPR & LightGCN & VBPR & MMGCN & SLMRec & LATTICE & FREEDOM & LGMRec & {DGAVE} & {LPIC} & \textbf{CRANE} \\
\hline
\multirow{4}{*}{Baby} & R@10 & 0.0357 & 0.0479 & 0.0423 & 0.0421 & 0.0521 & 0.0547 & 0.0627 & \underline{0.0644} & {0.0636} & {0.0634} & \textbf{0.0653} \\
                      & R@20 & 0.0575 & 0.0754 & 0.0663 & 0.0660 & 0.0772 & 0.0850 & 0.0992 & 0.1002 & {\underline{0.1009}} & {0.0977} & \textbf{0.1021} \\
                      & N@10 & 0.0192 & 0.0257 & 0.0223 & 0.0220 & 0.0289 & 0.0292 & 0.0330 & \underline{0.0349} & {0.0340} & {0.0337} & \textbf{0.0362} \\
                      & N@20 & 0.0249 & 0.0328 & 0.0284 & 0.0282 & 0.0354 & 0.0370 & 0.0424 & \underline{0.0440} & {0.0436} & {0.0422} & \textbf{0.0458} \\
\hline
\multirow{4}{*}{Sports} & R@10 & 0.0432 & 0.0569 & 0.0558 & 0.0401 & 0.0663 & 0.0620 & 0.0717 & 0.0720 & {\underline{0.0743}} & {0.0737} & \textbf{0.0746} \\
                        & R@20 & 0.0653 & 0.0864 & 0.0856 & 0.0636 & 0.0990 & 0.0953 & 0.1089 & 0.1068 & {\underline{0.1127}} & {0.1113} & \textbf{0.1129} \\
                        & N@10 & 0.0241 & 0.0311 & 0.0307 & 0.0209 & 0.0365 & 0.0335 & 0.0385 & 0.0390 & {\underline{0.0410}} & {0.0398} & \textbf{0.0412} \\
                        & N@20 & 0.0298 & 0.0387 & 0.0384 & 0.0270 & 0.0450 & 0.0421 & 0.0481 & 0.0480 & {\underline{0.0501}} & {0.0485} & \textbf{0.0503} \\
\hline
\multirow{4}{*}{Clothing} & R@10 & 0.0206 & 0.0361 & 0.0281 & 0.0227 & 0.0442 & 0.0492 & \underline{0.0629} & 0.0555 & {0.0619} & {0.0627} & \textbf{0.0642} \\
                          & R@20 & 0.0303 & 0.0544 & 0.0415 & 0.0361 & 0.0659 & 0.0733 & \underline{0.0941} & 0.0828 & {0.0917} & {0.0928} & \textbf{0.0956} \\
                          & N@10 & 0.0114 & 0.0197 & 0.0158 & 0.0120 & 0.0241 & 0.0268 & \underline{0.0341} & 0.0302 & {0.0336} & {0.0338} & \textbf{0.0352} \\
                          & N@20 & 0.0138 & 0.0243 & 0.0192 & 0.0154 & 0.0296 & 0.0330 & \underline{0.0420} & 0.0371 & {0.0412} & {0.0405} & \textbf{0.0431} \\
\hline
\multirow{4}{*}{{Electronics}} & {R@10} & {0.0372} & {0.0393} & {0.0293} & {0.0217} & {-} & {0.0364} & {0.0427} & {\underline{0.0440}} & {0.0432} & {-} & {\textbf{0.0449}} \\
                                          & {R@20} & {0.0557} & {0.0579} & {0.0458} & {0.0339} & {-} & {0.0551} & {0.0647} & {\underline{0.0661}} & {0.0657} & {-} & {\textbf{0.0678}} \\
                                          & {N@10} & {0.0208} & {0.0224} & {0.0159} & {0.0117} & {-} & {0.0207} & {0.0239} & {\underline{0.0244}} & {0.0242} & {-} & {\textbf{0.0251}} \\
                                          & {N@20} & {0.0256} & {0.0272} & {0.0202} & {0.0149} & {-} & {0.0261} & {0.0295} & {\underline{0.0304}} & {0.0301} & {-} & {\textbf{0.0390}} \\
\hline
\end{tabular*}
\vspace{15pt}
\end{table*}

\subsection{Overall Performance (RQ1)}

\Cref{tab:performance} reports the performance of CRANE and ten baselines across four datasets. The results consistently demonstrate the superiority of our proposed framework. We highlight five critical observations regarding the model's efficacy:

\begin{itemize}
    \item[--] \textbf{Necessity of Multimodal Signals.} Consistent with prior studies, multimodal models universally outperform traditional collaborative filtering methods. For instance, on the Clothing dataset, CRANE (Recall@20=0.0956) exceeds LightGCN (0.0544) by 75.6\%. This confirms that integrating visual and textual features is fundamental for capturing user preferences invisible in sparse interaction logs.

    \item[--] \textbf{Structural Advantage over Static Graphs.} Compared to graph-based multimodal baselines like FREEDOM (which freezes the item-item graph), CRANE achieves substantial margins (e.g., +2.9\% Recall@20 on Baby). This reveals the limitation of static pruning. In contrast, our dual-graph framework dynamically integrates recursive semantic alignment with collaborative propagation, allowing for a more flexible preference capture.

    \item[--] \textbf{Explicit Modeling vs. Implicit Alignment.}  Crucially, CRANE outperforms recent generative (DGAVE) and prompt-based (LPIC) approaches. While DGAVE utilizes VAEs for interpretability and LPIC leverages prompts for alignment, CRANE maintains a clear lead. This suggests that DGAVE’s variational inference may abstract away specific \textit{topological connectivity} crucial for ranking, while CRANE's explicit GCN backbone captures these structural signals more effectively.

    \item[--] \textbf{Robustness in Extreme Sparsity.} The advantage of CRANE is particularly pronounced on the Electronics dataset (99.99\% sparsity). In this challenging scenario where collaborative signals are fragmented, CRANE achieves the highest Recall@20 (0.0678). This validates that the homogeneous item-item graph acts as a critical \textit{semantic bridge}, propagating preferences between similar items even without direct co-interactions.

\end{itemize}

\subsection{Ablation Study of Core Modules}

To quantify the impact of CRANE's key components and verify the robustness of our design choices, we conduct a comprehensive ablation study on three diverse datasets (Baby, Sports, and Clothing). We evaluate six model variants to isolate the contributions of specific structural and functional modules. The variants are defined as follows:

\begin{itemize}
    \item \textbf{w/o Item Graph:} Removes the item-item semantic graph ($G_{II}$), relying solely on the user-item interaction graph ($G_{UI}$) for structural learning.
    \item \textbf{w/o RCA:} Disables the recursive mechanism in the Cross-Modal Attention module, employing only a single-layer attention for modality fusion ($R=1$).
    \item \textbf{w/o Attention:} Replaces the adaptive cross-modal attention mechanism with simple feature concatenation, removing the capability to model dynamic modality weights.
    \item \textbf{w/o GCN:} Removes the graph convolutional layers from both $G_{UI}$ and $G_{II}$, utilizing only raw interaction and modality features without high-order propagation.
    \item \textbf{w/o CL:} Removes the self-supervised contrastive alignment loss ($\mathcal{L}_{CL}$), optimizing the model solely with the recommendation objective ($\mathcal{L}_{BPR}$).
    \item \textbf{w/o Dual Fusion:} Disables the unified learning framework. Instead, it trains the collaborative branch (on $G_{UI}$) and semantic branch (on $G_{II}$) independently and employs a late fusion strategy for inference.
\end{itemize}

\begin{table*}[t]
  \centering
  \caption{Ablation Study of Core Modules across Three Datasets.}
  \label{tab:ablation}
  \renewcommand{\arraystretch}{1.1}
  \setlength{\tabcolsep}{3.5mm}
  
    { 
    \begin{tabular}{lcc|cc|cc}
      \toprule
      \multirow{2}{*}{\textbf{Method}} & \multicolumn{2}{c|}{\textbf{Baby}} & \multicolumn{2}{c|}{\textbf{Sports}} & \multicolumn{2}{c}{\textbf{Clothing}} \\
      \cline{2-7}
       & R@20 & N@20 & R@20 & N@20 & R@20 & N@20 \\
      \midrule
      w/o Item Graph   & 0.0965 & 0.0428 & 0.1072 & 0.0468 & 0.0912 & 0.0395 \\
      w/o RCA          & 0.0982 & 0.0439 & 0.1095 & 0.0481 & 0.0935 & 0.0412 \\
      w/o Attention    & 0.0975 & 0.0432 & 0.1088 & 0.0475 & 0.0928 & 0.0408 \\
      w/o GCN          & 0.0910 & 0.0395 & 0.1025 & 0.0442 & 0.0885 & 0.0380 \\
      w/o CL           & 0.0935 & 0.0408 & 0.1050 & 0.0455 & 0.0892 & 0.0390 \\
      w/o Dual Fusion  & 0.0958 & 0.0422 & 0.1065 & 0.0462 & 0.0905 & 0.0402 \\
      \midrule
      \textbf{CRANE (Ours)} & \textbf{0.1021} & \textbf{0.0458} & \textbf{0.1129} & \textbf{0.0503} & \textbf{0.0956} & \textbf{0.0431} \\
      \bottomrule
    \end{tabular}
  }
\end{table*}
\vspace{7pt}

The results in \Cref{tab:ablation} demonstrate that the full CRANE architecture consistently yields the highest performance. By comparing the full model with its variants, we derive the following insights:

\begin{itemize}
    \item[--] \textbf{Structural Dependency (GCN \& Item Graph).} The performance hierarchy \textit{Full Model} $>$ \textit{w/o Item Graph} $>$ \textit{w/o GCN} confirms that structural propagation is the backbone of our system. Specifically, removing the item graph leads to a notable degradation (e.g., -5.5\% Recall@20 on Baby), validating that the homogeneous semantic graph acts as a crucial bridge to propagate preferences among items with no co-interactions. The drastic drop in \textit{w/o GCN} further proves that topological connectivity is essential for alleviating data sparsity.

    \item[--] \textbf{Fusion Depth and Dynamics (RCA \& Attention).} The comparison between \textit{Full Model}, \textit{w/o RCA}, and \textit{w/o Attention} highlights the necessity of deep, adaptive fusion. Simple concatenation (\textit{w/o Attention}) fails to distinguish informative signals from noise, leading to suboptimal results. Furthermore, the performance gap between the recursive and single-layer settings (\textit{w/o RCA}) demonstrates that iterative refinement is key to uncovering high-order intra- and inter-modal dependencies that shallow fusion misses.

    \item[--] \textbf{Alignment of Views (CL \& Dual Fusion).} The results for \textit{w/o CL} and \textit{w/o Dual Fusion} confirm alignment is crucial. Since isolated processing (Late Fusion) is insufficient, the significant contrastive loss contribution mandates explicitly aligning the collaborative and semantic views in a shared latent space for mutual signal reinforcement.
\end{itemize}

Consequently, the ablation results empirically validate that CRANE's superior performance stems from the synergistic integration of its dual-graph architecture and recursive fusion mechanism. Each module plays a distinct yet indispensable role: the graphs provide the structural backbone, the RCA module refines the semantic alignment, and the contrastive objective ensures the robust unification of these views. Removing any component destroys this synergy, leading to the observed performance degradation.

\subsection{Evaluation of Multimodal Fusion Strategies}
\label{sec:fusion_evaluation}

To validate the efficacy of our RCA mechanism, we benchmark CRANE against several prevailing static fusion strategies. As discussed in the related work (\Cref{sec:multimodal-fusion}), multimodal fusion is generally categorized into \textit{early fusion} (feature-level) and \textit{late fusion} (decision-level). Accordingly, we define specific variants to represent these paradigms:

\begin{itemize}
    \item \textbf{Single-Modality Variants:}
    \begin{itemize}
        \item \textbf{CRANE-V:} Constructs the graph using exclusively visual features $\mathbf{X}^v$.
        \item \textbf{CRANE-T:} Constructs the graph using exclusively textual features $\mathbf{X}^t$.
    \end{itemize}
    
    \item \textbf{Early Fusion Variants (Feature-level):}
    \begin{itemize}
        \item \textbf{CRANE-C (Concatenation):} Uses feature concatenation, defined as $\mathbf{X}_{in} = [\mathbf{X}^v; \mathbf{X}^t]$.
        \item \textbf{CRANE-S (Summation):} Uses element-wise summation, defined as $\mathbf{X}_{in} = \mathbf{X}^v + \mathbf{X}^t$.
    \end{itemize}
    
    \item \textbf{Late Fusion Variant (Graph-level):}
    \begin{itemize}
        \item \textbf{CRANE-A (Average):} Computes the mean of independent visual ($\mathbf{S}^v$) and textual ($\mathbf{S}^t$) similarity matrices: $\mathbf{S}_{avg} = \frac{1}{2}(\mathbf{S}^v + \mathbf{S}^t)$.
    \end{itemize}
\end{itemize}

\Cref{fig:ablation-modality}   presents the comparative results. We summarize three critical observations regarding modality interaction:

\begin{figure}[!t]
  \centering
  \includegraphics[width=\linewidth]{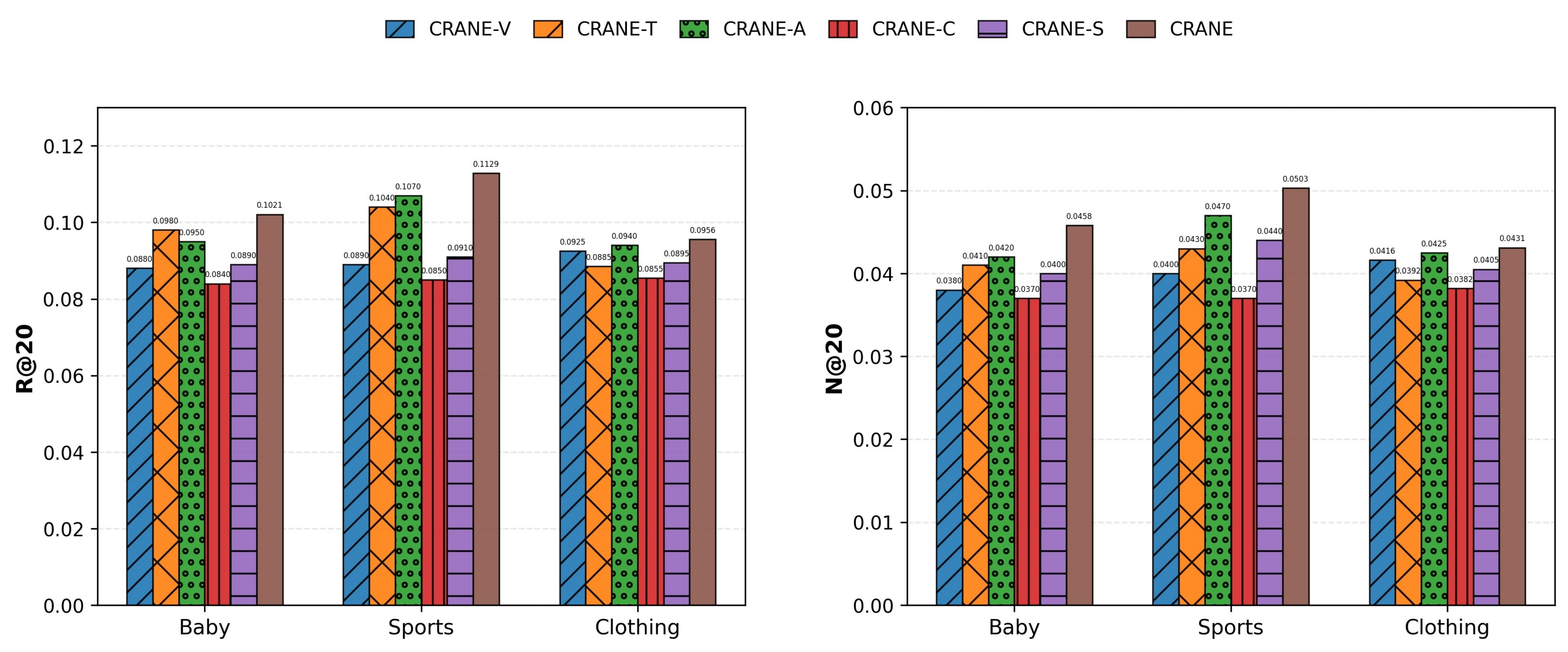}
  \caption{Evaluation of Multimodal Fusion Strategies}
  \label{fig:ablation-modality}
  \vspace{-6pt} 
\end{figure}

\begin{itemize}
    \item[--] \textbf{Inefficacy of Static Early Fusion.} Simple concatenation (CRANE-C) often fails to outperform single-modality baselines, particularly on the Clothing dataset (0.0855 vs. 0.0925 for Visual). This indicates that static fusion without adaptive alignment tends to introduce feature noise and mutual inhibition rather than synergy.
    
    \item[--] \textbf{Domain Sensitivity.} Performance varies significantly by domain—textual features dominate in Baby, while visual features are critical in Clothing. While Late Fusion (CRANE-A) offers a robust baseline by averaging these signals, it lacks the granularity to dynamically weigh the dominant modality for specific user-item pairs.
    
    \item[--] \textbf{Superiority of Recursive Alignment.} The full CRANE model consistently yields the highest performance across all datasets. This confirms that our RCA mechanism successfully captures deep intra- and inter-modal dependencies, offering a significant advantage over both shallow early fusion and static late fusion strategies.
\end{itemize}

In conclusion, the RCA mechanism is pivotal for effective fusion, enabling the discovery of deep inter-modal synergies that shallow methods fail to capture.

\subsection{Sensitivity Analysis of Key Hyperparameters}
\label{sec:sensitivity}
\begin{figure}[!t]
  \centering
  
  \begin{subfigure}[b]{0.8\linewidth} 
    \centering
    \setlength{\abovecaptionskip}{3pt} 
    
    \includegraphics[width=\linewidth]{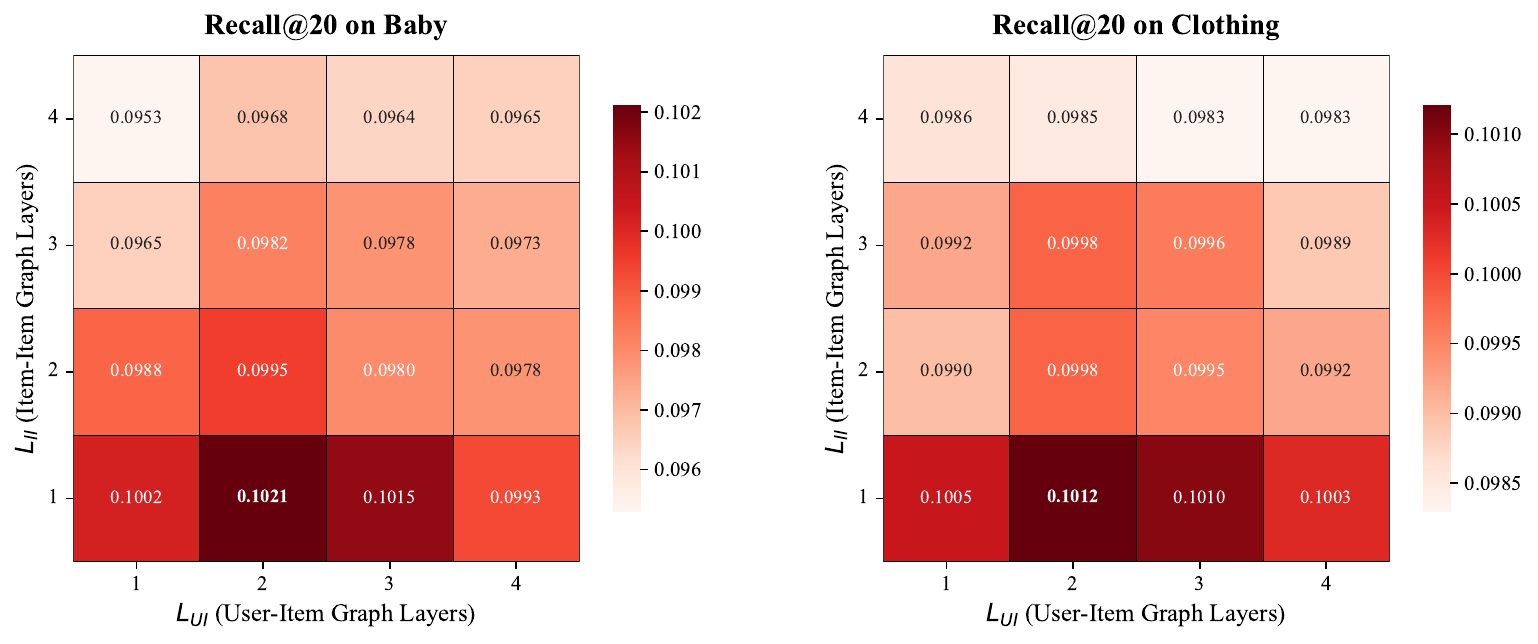}
    \caption{Impact of GCN Layer Depths ($L_{UI}$ vs. $L_{II}$)} 
    \label{fig:gcn_layers}
  \end{subfigure}
  
  \vspace{0.3cm} 
  
  \begin{subfigure}[b]{0.8\linewidth}
    \centering
    \setlength{\abovecaptionskip}{3pt} 
    
    \includegraphics[width=\linewidth]{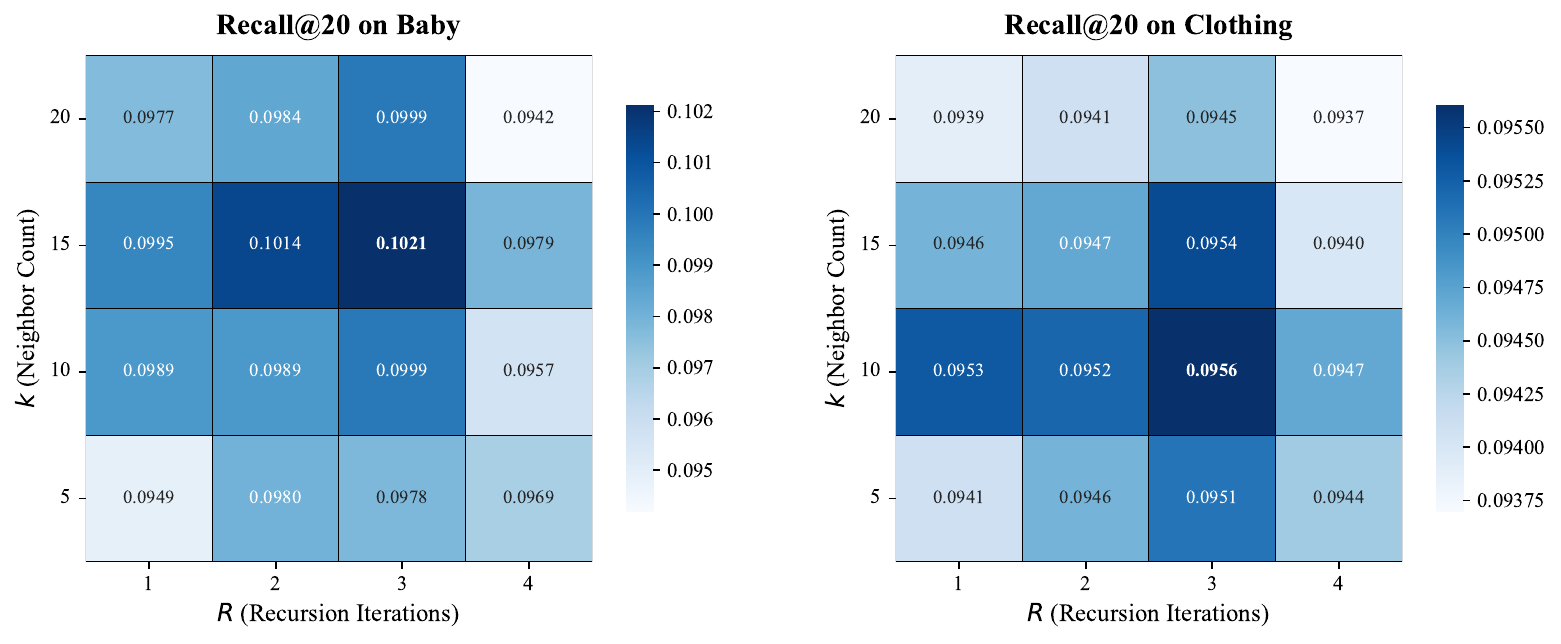}
    \caption{Impact of Recursion Depth ($R$) and Neighbor Count ($k$)}
    \label{fig:rk_params}
  \end{subfigure}
  
  \setlength{\abovecaptionskip}{8pt}
  \setlength{\belowcaptionskip}{0pt} 
  
\caption{Sensitivity analysis of key hyperparameters on. The darker color indicates better performance.}
\label{fig:sensitivity_analysis}
\end{figure}

The selection of critical hyperparameters—specifically the graph layer depths and the recursive attention parameters—is justified by theoretical constraints (e.g., over-smoothing) and empirical validation. We analyze two key sets of parameters, focusing on the Recall@20 metric, as visualized in  \Cref{fig:sensitivity_analysis}.

\label{sec:layer_sensitivity}

\subsubsection{Analysis of GCN Layer Depths ($L_{UI}$ and $L_{II}$). }
In our design, we set the depths of the User-Item Graph ($\mathcal{G}_{\text{UI}}$) and the Item-Item Graph ($\mathcal{G}_{\text{II}}$) to 2 and 1, respectively. This configuration is supported by both prior literature and our empirical hyperparameter analysis, as visualized in the heatmaps of \Cref{fig:gcn_layers}. The results indicate that the two graphs require different propagation depths due to their distinct topological characteristics: 

\begin{itemize}
    \item[--] \textbf{Necessity of Depth for Collaborative Propagation ($L_{UI}$). }The performance consistently peaks at $L_{UI}=2$ across both datasets (e.g., Recall@20 of 0.1021 on Baby). This confirms that a 2-layer architecture is essential for the bipartite interaction graph to capture the second-order connectivity (User $\to$ Item $\to$ User), which is the fundamental unit of collaborative filtering. Notably, increasing $L_{UI}$ to 3 maintains competitive performance (0.1015), indicating the model's robustness to collaborative depth.

    \item[--] \textbf{Sensitivity to Semantic Over-smoothing ($L_{II}$).} 
Conversely, the model exhibits high sensitivity to the depth of the semantic graph. Increasing $L_{II}$ from the optimal value of 1 to 2 causes an immediate performance drop (from 0.1021 to 0.0995 on Baby). This sharp decline is attributed to the dense nature of the k-NN-based semantic graph ($k=15$). Unlike the sparse interaction graph, propagating information beyond 1-hop in a dense similarity graph rapidly leads to \textit{over-smoothing}, where item representations become indistinguishable due to the aggregation of distant, noisy semantic neighbors.
\end{itemize}

Consequently, the $(2, 1)$ configuration is not a heuristic choice but a statistically optimal balance between capturing long-range collaborative signals and preserving precise local semantic features.

\subsubsection{Analysis of Recursion Depth ($R$) and Neighbor Count ($k$). }
We investigate the impact of the RCA depth ($R$) and the graph density parameter ($k$) for the item-item graph, as visualized in Figure~\ref{fig:rk_params}. The results reveal two distinct trends regarding the model's sensitivity to computational depth and topological density:

\begin{itemize}
    \item[--] \textbf{Recursive Mechanism Efficacy ($R$).} Increasing the recursion depth yields observable gains, peaking at $R=3$ (e.g., Recall@20 rises from 0.0977 to \textbf{0.1021} on Baby with $k=15$). This confirms that \textit{iterative refinement} is essential for capturing high-order cross-modal dependencies. However, performance declines at $R=4$, indicating a potential risk of feature over-refinement.
    
    \item[--] \textbf{Optimal Graph Density ($k$).} The model performs optimally at moderate density levels ($k=15$ for Baby, $k=10$ for Clothing). Extremely sparse graphs ($k=5$) limit semantic message propagation, while excessive density ($k=20$) introduces noise, effectively diluting the valid semantic signals.
\end{itemize}

Based on these observations, we adopt $R=3$ across datasets and tune $k$ individually, ensuring an optimal balance between semantic depth and structural density.

\subsection{User Profile Aggregation Strategy Evaluation}
\label{sec:user-profile}

To validate the rationale behind employing element-wise summation for user profile construction, we conduct a comprehensive ablation study on the Clothing dataset. We compare our approach against three alternative strategies:

\begin{itemize}

    \item \textbf{Average Pooling:} Normalizes the sum , defined as $\mathbf{x}_u^m = \frac{1}{|\mathcal{N}(u)|} \sum_{i \in \mathcal{N}(u)} \mathbf{x}_i^m$.
    \item \textbf{Max Pooling:} Selects the most salient feature dimensions via $\mathbf{x}_u^m = \max_{i \in \mathcal{N}(u)} (\mathbf{x}_i^m)$.
    \item \textbf{Attention-Weighted :} Learns adaptive importance weights $\alpha_{ui}$ via $\mathbf{x}_u^m = \sum_{i \in \mathcal{N}(u)} \alpha_{ui} \mathbf{x}_i^m$.
    \item \textbf{Summation (Ours):} Accumulates total signal intensity via $\mathbf{x}_u^m = \sum_{i \in \mathcal{N}(u)} \mathbf{x}_i^m$.
\end{itemize}

The experimental results are summarized in \Cref{tab:tab-user-profile} below:

\begin{table}[h]
\centering

\setcounter{table}{5}

\refstepcounter{table}

\caption*{\centering \large Table \thetable. Performance and Efficiency of Aggregation Strategies on $\text{Clothing}$ Dataset}

\label{tab:tab-user-profile}
\vspace{3pt}
\resizebox{0.95\linewidth}{!}{
\begin{tabular}{cccccc}
\toprule
\textbf{Aggregation Strategy} & \textbf{Recall@10} & \textbf{Recall@20} & \textbf{NDCG@10} & \textbf{NDCG@20} & \textbf{Time (s)} \\
\midrule
Average Pooling & 0.0640 & 0.0952 & 0.0350 & 0.0428 & 4.77\\
Max Pooling & 0.0637 & 0.0948 & 0.0348 & 0.0425 & 4.72\\
Attention Aggregation & 0.0641 & 0.0954 & \textbf{0.0353} & 0.0431 & 4.98\\
\textbf{Summation (Ours)} & \textbf{0.0642} & \textbf{0.0956} & \textbf{0.0353} & \textbf{0.0432} & 4.76\\
\bottomrule
\end{tabular}
}
\end{table}

Based on these results, our comparative analysis yields three critical insights regarding the trade-off between complexity and effectiveness:

\begin{itemize}
  
    \item[--] The summation strategy consistently yields the highest retrieval performance across metrics (e.g., Recall@20 of 0.0956). This superiority stems from its ability to preserve the natural intensity of user preferences inherent in interaction frequency. In contrast, Average Pooling suffers from signal dilution by normalizing diverse interaction counts, effectively suppressing the confidence of active users. Similarly, Max Pooling selects only dominant features, failing to capture the full spectrum of a user's multi-faceted interests.

    \item[--] The attention mechanism introduces noticeable computational overhead (rising to 4.98s per epoch) due to the dense matrix operations required for dynamic weight calculation. Furthermore, the introduction of additional trainable parameters appears to induce overfitting, particularly on sparse interaction data. Instead of refining semantic alignment, the attention mechanism tends to inadvertently amplify noise or learn spurious correlations when supervision signals are scarce.

    \item[--] Regarding the overall trade-off, summation maintains a high inference speed (4.76s), comparable to the simplest baselines like Max Pooling, yet achieves superior accuracy. This empirical evidence confirms that preserving cumulative signal intensity provides the most robust and efficient solution for implicit feedback. It effectively avoids the latency and optimization stability penalties associated with complex learnable mechanisms, offering a streamlined architecture suitable for large-scale applications.
\end{itemize}

Consequently, we adopt Summation as the optimal strategy. This choice prioritizes cumulative signal preservation over complex parameterization, delivering representations that are as expressive as learnable mechanisms while maintaining maximum computational efficiency and model simplicity.

\begin{figure}[t]
  \centering
  \includegraphics[width=0.75\linewidth, trim=10 10 10 10, clip]{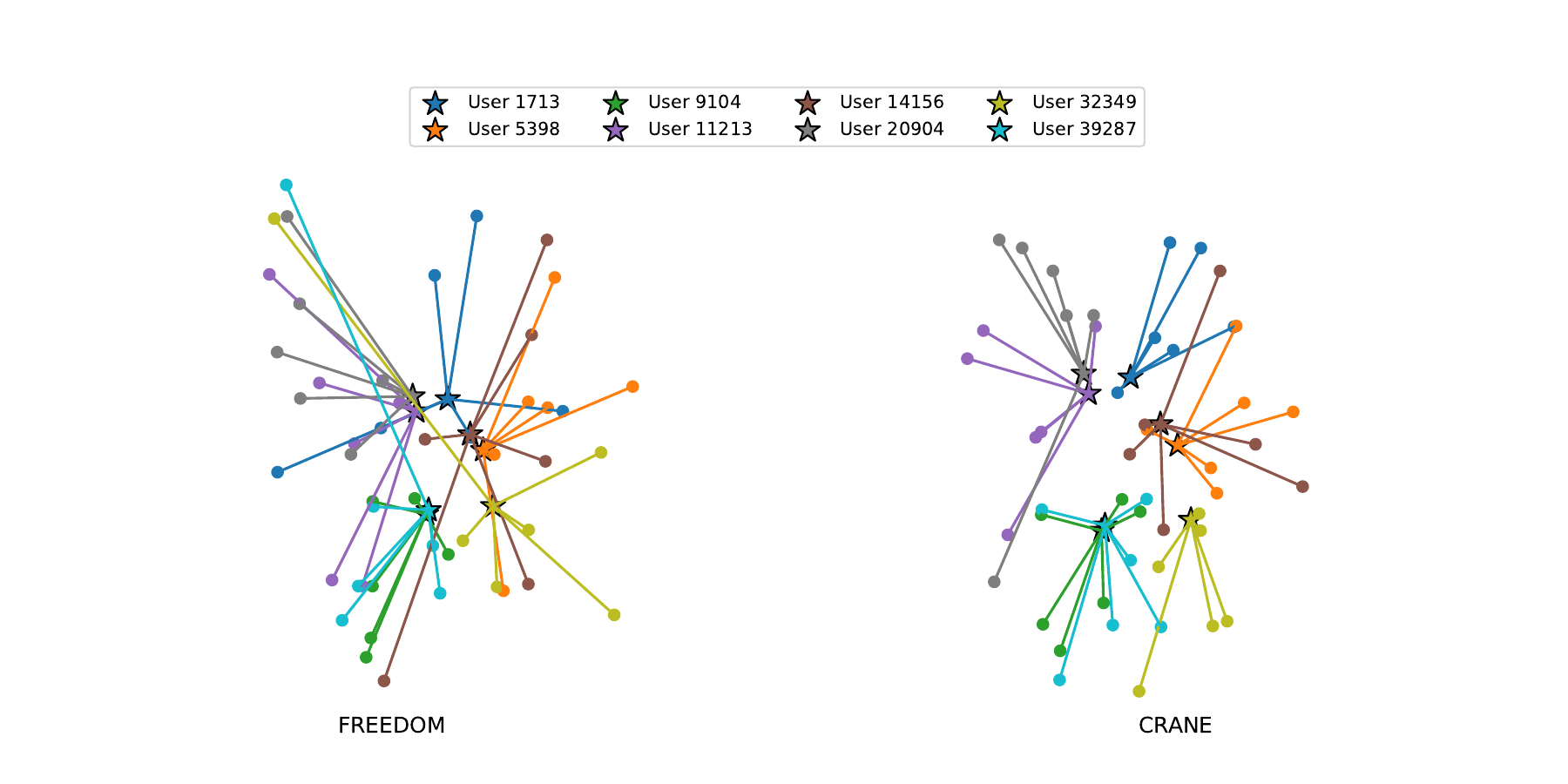}
  \caption{t-SNE  of user ($\star$) and item embeddings. CRANE shows markedly tighter user-item alignment.}
  \label{fig:Visualization}
  \vspace{-1pt} 
\end{figure}

\subsection{Visualization Analysis }

To qualitatively assess preference capture, \Cref{fig:Visualization} visualizes the t-SNE embeddings of eight randomly sampled users ($\star$) and their interacted items (points) from the Baby dataset. In contrast to the baseline FREEDOM, which exhibits a dispersed distribution with blurred boundaries between user groups, CRANE generates significantly tighter and more distinct clusters. This superior structural alignment provides intuitive evidence that our dual-graph architecture and RCA mechanism effectively filter modality noise to enforce high intra-user compactness. Consequently, the model demonstrates enhanced discriminative power, successfully projecting diverse multimodal inputs into a cohesive, user-centric latent space.

\section{Efficiency and Scalability Analysis}
\label{sec:efficiency}

\subsection{Theoretical Complexity}
\label{sec:complexity}

We analyze the asymptotic time complexity of CRANE, focusing on the trade-off between structural propagation efficiency and semantic fusion depth. The computational process is divided into two distinct phases. First, the structural learning phase utilizes heterogeneous graph convolution to capture local collaborative signals. Since the message passing on both the user-item graph $\mathcal{G}_{UI}$ and the $k$-NN sparsified item graph $\mathcal{G}_{II}$ is implemented via sparse matrix-vector multiplications (SpMV), the computational cost is strictly proportional to the edge density rather than the matrix dimension. This results in linear complexity terms of $\mathcal{O}(L_{UI}\|\mathbf{A}\|_0 d)$ and $\mathcal{O}(L_{II} N k d)$, ensuring scalability over sparse interaction data. Second, the rca module aims to capture global intra- and inter-modal dependencies. Unlike the local aggregation in GCNs, this phase requires dense matrix multiplications to compute the pairwise cross-modal correlation matrix $\mathbf{C}^m$ in each of the $R$ iterations. This global alignment introduces a quadratic cost relative to the number of entities. Consequently, the total theoretical complexity is derived as:
\begin{equation}
    \mathcal{T} = \mathcal{O}\left( (L_{UI}\|\mathbf{A}\|_0 + L_{II} N k + R N^2) \cdot d \right)
\end{equation}
While asymptotically quadratic ($\mathcal{O}(N^2)$), this theoretical cost does not dominate execution time on typical datasets. The subsequent scalability analysis indicates that CRANE follows a near-linear runtime trajectory on the evaluated large-scale benchmarks.

\subsection{Scalability Validation and Practical Limits}

\begin{figure*}[t]
  \centering
  \begin{subfigure}[b]{0.48\linewidth}
    \centering
    \includegraphics[width=\linewidth]{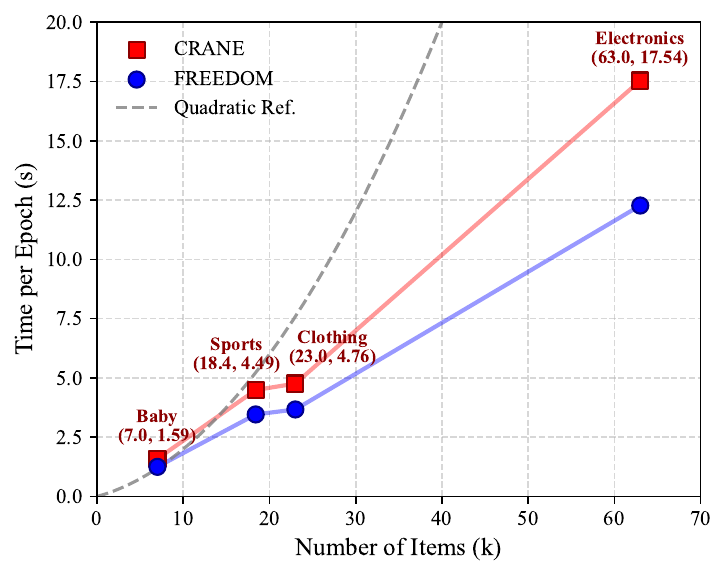}
    \caption{Runtime Scaling with Item Quantity}
    \label{fig:runtime}
  \end{subfigure}
  \hfill 
  \begin{subfigure}[b]{0.48\linewidth}
    \centering
    \includegraphics[width=\linewidth]{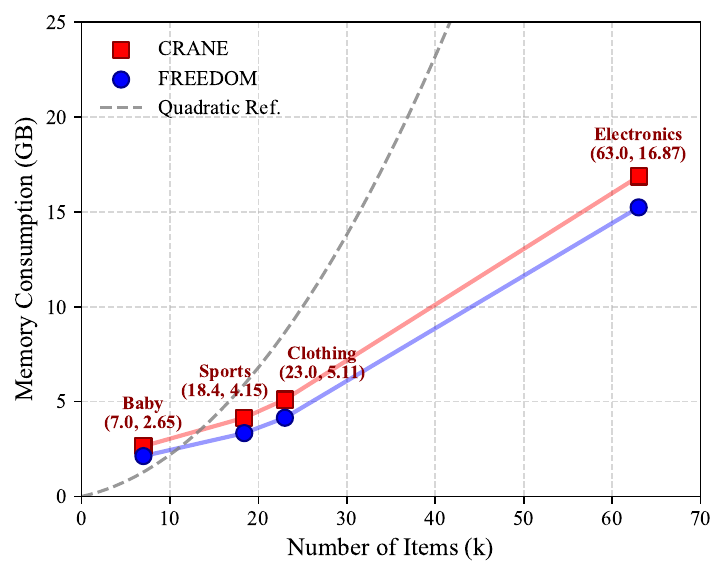}
    \caption{Memory Consumption Comparison}
    \label{fig:memory}
  \end{subfigure}

  \caption{Scalability analysis of CRANE compared to FREEDOM. (a) Runtime scaling showing trends relative to linear and quadratic references. (b) Memory consumption comparison across datasets.}
  \label{fig:scalability}
\end{figure*}

To address concerns regarding the quadratic attention term, we benchmark CRANE against the highly efficient linear baseline FREEDOM ~\cite{DBLP:conf/mm/ZhouS23} on the large-scale \textit{Electronics} dataset ($N > 63k$). \Cref{fig:scalability} visualizes the comparative time per epoch and memory consumption.

\textbf{Empirical Scaling Analysis.} Despite the theoretical $\mathcal{O}(N^2)$ complexity, empirical results in \Cref{fig:runtime}  indicate that CRANE exhibits a runtime scaling trajectory comparable to the linear baseline. On \textit{Electronics}, CRANE records 17.54s per epoch, presenting a manageable overhead. This suggests that the computational cost is largely influenced by memory-bound sparse graph convolutions, mitigating the latency impact of dense attention at this scale. Regarding memory, \Cref{fig:memory} (16.87 GB on \textit{Electronics}) displays a growth pattern approximately proportional to dataset size, attributed to the adoption of Compressed Sparse Row (CSR) formats which facilitate efficient resource management.

\textbf{Practical Limits and Future Optimizations.} While efficient on standard benchmarks, we acknowledge that the $\mathcal{O}(N^2)$ term implies a theoretical bottleneck for ultra-large-scale applications (e.g., $N > 10^6$). In such scenarios, the memory requirement for a dense correlation matrix may exceed standard GPU capacities. Nevertheless, the architecture offers potential for algorithmic optimization. Future investigations might explore strategies such as block-sparse masks or locality-sensitive hashing (LSH) to prune negligible correlations. These techniques provide a theoretical basis for reducing complexity towards $\mathcal{O}(N \log N)$, aiming to preserve the capture of core cross-modal dependencies.

\subsection{Convergence and Learning Dynamics}
\label{sec:convergence}

\begin{figure*}[!t]
 \centering
 \includegraphics[width=\linewidth]{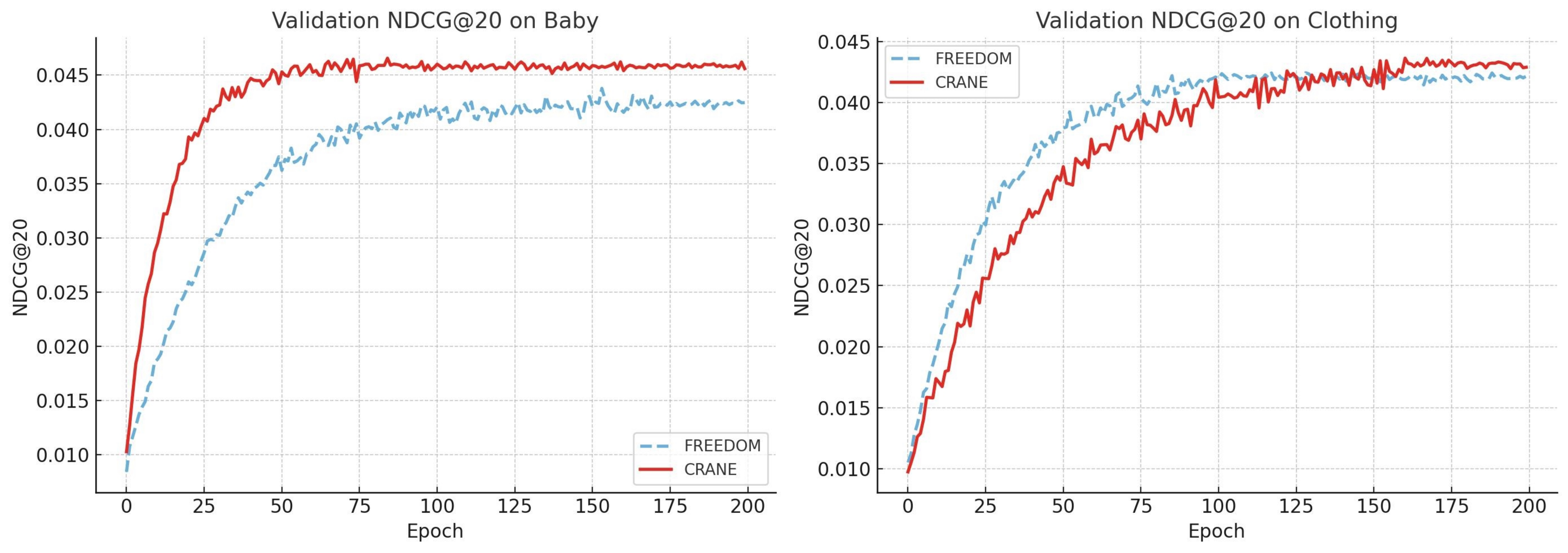}
 \caption{ Efficiency Comparison of CRANE and FREEDOM on Baby and Clothing with each epoch.
}
 \label{fig:convergence}
\end{figure*}
Beyond per-epoch costs, we evaluate the time-to-accuracy efficiency by analyzing the validation NDCG@20 curves in \Cref{fig:convergence}. The results highlight distinct behaviors across dataset scales. On the sparse Baby dataset, CRANE demonstrates superior convergence speed, reaching optimal performance within just 25 epochs—approximately $4\times$ faster than the baseline. Conversely, on the larger Clothing dataset, the advantage shifts to a higher performance ceiling: while FREEDOM suffers from early saturation around epoch 75, CRANE sustains steady improvement to reach a peak of 0.0431, effectively breaking the baseline's bottleneck.

These observations underscore the model's dynamic capability to adapt its learning trajectory based on data complexity. The rapid uptake on sparse data suggests that the RCA mechanism can efficiently extract high-order cross-modal signals from limited interactions, making it highly resource-efficient for rapid prototyping. Meanwhile, the sustained improvement on complex data demonstrates that the dual-graph architecture effectively aligns collaborative and semantic spaces to avoid local optima. In summary, CRANE offers a versatile profile: it not only lowers the training barrier for sparse scenarios but also possesses the potential for robust, sustained learning on larger-scale tasks.

\section{Conclusion}
\label{sec:conclusion}

In this paper, we introduced CRANE, a novel framework designed to overcome the persistent limitations of shallow modality fusion and asymmetric representation in multimodal recommendation. By synergizing a symmetric dual-graph architecture with a recursive cross-modal attention (RCA) mechanism, our model successfully captures high-order intra- and inter-modal dependencies while explicitly constructing semantic profiles for both users and items. Extensive experiments across four diverse datasets validate that CRANE consistently outperforms state-of-the-art baselines. Notably, empirical scalability analysis validates that CRANE maintains high practical efficiency on large-scale benchmarks, offering a superior trade-off between model expressiveness and computational cost. Moving forward, we aim to optimize the framework for dynamic graph scenarios to handle real-time streaming data. 

\begin{acks}
This work was supported by Beijing Natural Science Foundation (JQ24019). This work was supported in part by the National Natural Science Foundation of China (No. 62576047, 52572349). This work is being Supported by the Open Fund of the Key Laboratory for Civil Aviation Collaborative Air Traffic Management Technology and Applications (No. 2025-001) and sponsored by SMP-Z Large Model Fund (No. CIPS-SMP20250313).

\clearpage

\end{acks}

\bibliographystyle{ACM-Reference-Format}
\bibliography{sample-base}

\end{document}